\documentstyle[12pt]{article}

\columnwidth\textwidth

\begin{document}

\newcommand{\reell}{{\kern+.25em\sf{R}\kern-.78em\sf{I}\kern+.78em\kern-.25em}}
\newcommand{\komplex}{{\sf{C}\kern-.46em\sf{I}\kern+.46em\kern-.25em}}
\newcommand{\posganz}{{\kern+.25em\sf{N}\kern-.86em\sf{I}\kern+.86em\kern-.25em}}
\newcommand{\ganz}{{\kern+.25em\sf{Z}\kern-.78em\sf{Z}\kern+.78em\kern-.65em}}
\newcommand{\Hamilton}{{\kern+.25em\sf{H}\kern-.86em\sf{I}\kern+.86em\kern-.25em}}
\newcommand{\Cayley}{{\sf{O}\kern-.56em\sf{I}\kern+.56em\kern-.25em}}
\newcommand{\unit}{{\sf{1}\kern-.18em\sf{I}\kern+.18em\kern-.18em}}
\newcommand{\opunit}{{\sf{1}\kern-.29em\sf{1}\kern+.29em\kern-.33em}}
\newcounter{blabla}
\newcommand{\be}{\begin{equation}}
\newcommand{\ee}{\end{equation}}
\newcommand{\bdm}{\begin{displaymath}}
\newcommand{\edm}{\end{displaymath}}
\newcommand{\beann}{\begin{eqnarray*}}
\newcommand{\eeann}{\end{eqnarray*}}
\newcommand{\bea}{\begin{eqnarray}}
\newcommand{\eea}{\end{eqnarray}}

\newcommand{\nn}{\nonumber}

\begin{titlepage}

\def\mytoday#1{{ } \ifcase\month \or
 January\or February\or March\or April\or May\or June\or
 July\or August\or September\or October\or November\or December\fi
 \space \number\year}
\noindent
\hspace*{11cm} BUTP--94/22\\
\vspace*{1cm}
\begin{center}
{\LARGE Group Quantization of Parametrized Systems
\\II. Pasting Hilbert spaces}

\vspace{2cm}

P. H\'{a}j\'{\i}\v{c}ek, A Higuchi and J. Tolar\footnote{Present address:
Department of Physics and
Doppler Institute, Faculty of Nuclear Sciences and Physical
Engineering, Czech Technical University, B\v rehov\'a 7,
CZ--115 19 Prague 1, Czech Republic
(e-mail: tolar@br.fjfi.cvut.cz).} \\
Institute for Theoretical Physics \\
University of Bern \\
Sidlerstrasse 5, CH-3012 Bern, Switzerland
\\ \vspace{1.5cm}

December 1994 \\ \vspace*{1cm}

\nopagebreak[4]

\begin{abstract}
The method of group quantization described in the preceeding paper I is
extended so that it becomes applicable to some parametrized systems that do not
admit a global transversal surface. A simple completely solvable toy system is
studied that admits a pair of maximal transversal surfaces intersecting all
orbits. The corresponding two quantum mechanics are constructed. The similarity
of the canonical group actions in the classical phase spaces on the one hand
and in the quantum Hilbert spaces on the other hand suggests how the two
Hilbert spaces are to be pasted together. The resulting quantum theory is
checked to be equivalent to that constructed directly by means of Dirac's
operator constraint method. The complete system of partial Hamiltonians for any
of the two transversal surfaces is chosen and the quantum Schr\"{o}dinger or
Heisenberg pictures of time evolution are constructed.

\end{abstract}

\end{center}

\end{titlepage}

\section{Introduction}

Quantization of first-class parametrized systems, like general relativity or
string theory, is an outstanding problem in theoretical physics. In ref.
\cite{1}, which will be referred to as I in this paper, a theory has been
worked out for parametrized systems of finite dimensions in which a crucial
role is played by gauge invariant objects like algebras of perennials or groups
of symmetries; the theory combines ideas on the forms of relativistic dynamics
\cite{2} and on algebraic \cite{3} and group \cite{4} quantization. The problem
of time, which is characteristic of the quantization of parametrized systems,
emerges in the form of ``frozen dynamics problem'' \cite{5}. The proposal in I
of how this problem can be solved goes back to Dirac's notions of transversal
surface and Hamiltonian \cite{2}. A transversal surface represents a particular
gauge, and a system of Hamiltonians defines a family of time levels in
classical solutions. It is shown in refs. \cite{1} and \cite{6} that different
choices of transversal surfaces and Hamiltonians lead to the same quantum
theories in the sense that the predictions for one and the same measurement is
independent of these choices. In other words, the so-called ``multiple choice
problem'' \cite{5}, which is typical of quantization methods that have to
specify a choice of time foliation, does not afflict our theory. In order to
obtain these results, however, the existence of a so-called global transversal
surface was assumed. Thus, we still have to confront the ``global time''
\cite{5} and ``Gribov problems'': There are many parametrized systems which do
not admit a global transversal surface.

As yet we are not able to give a general solution to this difficult problem;
the present paper is just a small exercise: we limit ourselves to a very simple
and completely solvable system for which there is no global transversal surface
and we describe some method of solution to the problem at least for this
system. The idea is, roughly speaking, taken over from the theory of
differentiable manifolds: if there is no global coordinate system, then there
will be coordinate patches; one can calculate within each patch and transform
the results between the patches where they overlap. Similarly, one can choose
a system of transversal surfaces which together intersect all gauge orbits and
which are maximal in the sense of paper I. Each of these surfaces defines a
reduced system of its own, for which quantum theories can be constructed. Then
the quantum theories corresponding to different transversal surfaces are
``pasted together'' in a self-consistent way so that a single quantum theory
results. In particular, the Hilbert spaces are pasted together so that again a
Hilbert space is obtained (and not, say, the more general Hilbert manifold).
The pasting method is based on the property of the groups of symmetries that
they ``remember '' the global structure of the transversal surfaces.

The plan of the paper is as follows. In Sec.\ 2, we define the classical model
and study its constraint surface, the topology of the orbits in the constraint
surface that is relevant to the nonexistence of a global transversal surface,
and
list some perennials. In Sec.\ 3,
a particular class of symmetries of the system
is considered, namely the transformations which act linearly in the extended
phase space. All continuous linear symmetries from the group $SO_2(2,1)$ (the
subscript ``2'' denotes the double covering); in addition, two discrete
symmetries, $\sigma$ and $\tau$ are found, which may be related to the parity
and time reversal in the 3-dimensional ``Minkowski'' space of the fundamental
representation of $SO(2,1)$. Then, we choose the group $O^+_2(2,1)$ (this is
the double covering of the orthochronous Lorentz group in three dimensions; it
includes the parity $\sigma$) as the first-class canonical group of the system
(cf. I). The action of $O^+_2(2,1)$ on the constraint hypersurface is then
explicitly calculated. In Sec.\ 4, we choose a pair $\Gamma_1, \Gamma_2$ of
maximal transversal surfaces, find their symplectic structure, determine their
overlapping and project the action of the group $O^+_2(2,1)$ to them. It turns
out that due to the inclusion of $\sigma$, the group acts transitively on both
transversal surfaces, and has a subgroup, namely $SO_2(2,1)$, which acts
transitively just within their overlapping subsets. In Sec.\ 5, we choose a
representation of the symmetry group. The method of constructing the
representation is just the canonical quantization of the reduced systems
associated with each transversal surface. In this way, one obtains a reducible
representation which decomposes into two irreducible ones. Only one of these,
however, is faithful and so the choice, at least between these two obtained by
canonical quantization, is easy. Moreover, the chosen representation of
$O^+_2(2,1)$ decomposes into two irreducible representations of its subgroup
$SO_2(2,1)$. Then, with each (classical) phase space $\Gamma_i, i=1,2$, we
associate a (quantum) Hilbert space ${\cal H}_i$ carrying the above
representation of $O^+_2(2,1)$. ${\cal H}_i$ has two subspaces, ${\cal
H}^\pm_i$, each of which carries an irreducible representation of $SO_2(2,1)$.
This is analogous to $\Gamma_i$ having two open submanifolds, $\Gamma^\pm_i$,
in which $\Gamma_1$ and $\Gamma_2$ overlap, and within which $SO_2(2,1)$ acts
transitively. This analogy between the action of these groups in the classical
and quantum theory suggests the construction of quantum mechanics that is
performed in Sec.\ 6.
There, we identify the subspace ${\cal H}^+_1$ with ${\cal
H}^+_2$ and ${\cal H}^-_1$ with ${\cal H}^-_2$ using the unitary
transformations
\bdm
U_\pm \; : \; {\cal H}^\pm_1 \rightarrow {\cal H}^\pm_2
\edm
that map the corresponding equivalent representations of $SO_2(2,1)$ into each
other. These ``pasting maps'' $U_\pm$ are determined up to phase factors
(Schur
lemma!). By the requirement that the parity acts on the Hilbert space obtained
by the identification, the relative phase factor is fixed. Sec.\ 7 contains a
proof that the pasting construction yields a quantum theory that is equivalent
to the quantum theory of the same system obtained by the
Wheeler-DeWitt-equation method -- a method which dispenses of transversal
surfaces. Thus, certain ``convergence'' of methods is demonstrated. Finally, in
Sec.\ 8, a system of Hamiltonians is chosen, the corresponding family of time
levels is shown to be complete and the Schr\"{o}dinger or Heisenberg picture of
time evolution is described in detail.

\section{The model}

In this section, we describe the chosen simple model.
The (extended) phase space $\Gamma$ is $(\reell^{4}, \Omega)$.
In global Cartesian coordinates $q_1$, $q_2$, $p_1$, $p_2$
the symplectic form is
$\Omega = dp_{1}\wedge dq_{1} + dp_{2}\wedge dq_{2}$. The definition of
our parametrized system is finished by specifying the constraint:
\begin{equation}
C = \frac{1}{2} (p_{1}^{2} - p_{2}^{2} - q_{1}^{2}
                  + q_{2}^{2}) = 0.    \label{0.}
\end{equation}

The model has
a purely quadratic constraint $C$. Thus, the equations
of motion are linear, with constant coefficients, and
the system is completely solvable. Still, the space
of $C$-orbits (classical solutions) is non-Hausdorff
and there are three topologically different kinds of
orbits: free orbits, which definitely leave any compact subset
of $\Gamma$ in both directions, orbits with a limit point
in the future or the past, and one critical point. As proposed in I, we remove
the critical point from the phase space. Thus, from now on $\Gamma$ stands for
$\reell \setminus \{(0,0,0,0)\}$.

\subsection{Constraint hypersurface}

Let  $\tilde{\Gamma}$ denote the constraint hypersurface
defined by  eq.\ (\ref{0.}). Introduce coordinates $r$, $\alpha$,
$\beta$ on  $\tilde{\Gamma}$ by
\begin{eqnarray*}
q_1  = -r \cos \beta,  \quad p_1 = r \cos \alpha, \\
q_2 = r \sin \alpha, \qquad  p_2  = r \sin \beta.
\end{eqnarray*}
For $r > 0$, $\alpha , \beta \in [0, 2\pi )$, we obtain
all points of  $\tilde{\Gamma}$.
The topology is
\bdm
\tilde{\Gamma} \; = S^{1} \times S^{1}
      \times {\reell}_{+}.
\edm

To visualize the orbits and find the symmetries,
a canonical transformation to the
following set of canonical coordinates is useful:
\begin{eqnarray*}
Q_{1}& = &\frac{1}{\sqrt{2}} (q_{1} + p_{1}), \quad
P_{1} = \frac{1}{\sqrt{2}} (-q_{1} +  p_{1}), \\
Q_{2} &= &\frac{1}{\sqrt{2}} (q_{2} - p_{2}), \qquad
P_{2} = \frac{1}{\sqrt{2}} (q_{2} +  p_{2}).
\end{eqnarray*}
In the new coordinates, the constraint reads
\bdm
C = Q_{1} P_{1} + Q_{2} P_{2}.
\edm

Any point of $\Gamma$ is determined by two vectors
\bdm
 Q = (Q_1,Q_2), \qquad  P = (P_1,P_2),
\edm
which can be pictured as vectors in the $Q$-
and $P$-planes, respectively.
The constraint $C=0$ just means that they are
`orthogonal'. This leads us to the third
coordinate system on  $\tilde{\Gamma}$: we define
\begin{eqnarray}
m(\psi) & = & (- \sin \frac{\psi}{2},
              \cos \frac{\psi}{2}),  \label{11.}\\
n(\psi) & =  & (\cos \frac{\psi}{2},
               \sin \frac{\psi}{2}),
\end{eqnarray}
and set
\begin{equation}
Q =y\,m(\psi) , \quad P =x\,n(\psi).    \label{12.}
\end{equation}
Clearly, the point of $\Gamma$ determined by $\psi$,
$x$, $y$ with $\psi \in [0,4\pi)$, $x,y \in \reell$, lies
in  $\tilde{\Gamma}$; however, each point of
 $\tilde{\Gamma}$ is obtained twice in this way, since
$(\psi,x,y)$ and $(\psi + 2\pi,-x,-y)$ determine the
same point.

In terms of $r$, $\alpha$, $\beta$ we have
\begin{eqnarray}
Q_1 & = & - \sqrt{2}\, r \sin \frac{\alpha + \beta}{2}
             \sin \frac{\alpha - \beta}{2} ,\label{14.}\\
Q_2  &= & \sqrt{2}\, r \cos \frac{\alpha + \beta}{2}
                \sin \frac{\alpha - \beta}{2} , \\
P_1  &= & \sqrt{2}\, r \cos \frac{\alpha + \beta}{2}
                \cos \frac{\alpha - \beta}{2} ,\\
P_2 &= & \sqrt{2}\, r \sin \frac{\alpha + \beta}{2}
            \cos \frac{\alpha - \beta}{2} ,\label{15.}
\end{eqnarray}
hence
\bdm
\psi = \alpha + \beta, \;\;
x = \sqrt{2}\, r \cos \frac{\alpha - \beta}{2}, \;\;
y = \sqrt{2}\, r \sin \frac{\alpha - \beta}{2}.
\edm

\subsection{Orbits}
The orbits are solutions to the system of
differential equations
\bdm
\dot Q = \{Q,C\} = Q, \quad
\dot P = \{P,C\} = P,
\edm
which is easy to integrate:
\bdm
Q(t) =Q(0)\ e^t, \quad P(t) =P(0)\ e^{-t}.
\edm
This implies immediately
\bdm
\psi (t) = \psi (0), \; x(t) = x(0)\  e^{-t},\;
y(t) = y(0)\ e^t,
\edm
hence $\psi$ and the product $xy$ is {\it constant along orbits}.

There are two disconnected sets of
{\it `free'} orbits (they leave definitely any compact
set for sufficiently large or small $t$):
\begin{enumerate}
\item[I.] $xy > 0$, i.e. $x>0$, $y>0$,
      $\psi \in [0,4\pi)$ \\
       (identical with  $x<0$, $y<0$,
         $\psi \in [0,4\pi)$ ),
\end{enumerate}
and
\begin{enumerate}
\item[II.] $xy < 0$, i.e. $x>0$, $y<0$,
      $\psi \in [0,4\pi)$ \\
       (identical with  $x<0$, $y>0$,
         $\psi \in [0,4\pi)$ ).
\end{enumerate}
They are separated by two-dimensional boundaries
formed by {\it `imprisoned'} orbits
\begin{enumerate}
\item[III.] $x=0$, $\psi \in [0,4\pi)$,
\end{enumerate}
and
\begin{enumerate}
\item[IV.] $y=0$, $\psi \in [0,4\pi)$.
\end{enumerate}
Observe that the space of orbits
 `$\tilde{\Gamma}$/orbits'
is not Hausdorff and that the orbits with $x=0$
are not `separable' from those with $y=0$.
There will be, therefore, no global transversal
surface \cite{7}.

\subsection{Perennials}
We have seen that
the variables $Q_{i}$, $P_{i}$ satisfy the equations
\bdm
\{Q_{i},C\} = Q_{i}, \quad  \{P_{i},C\} = -P_{i}.
\edm
Thus we can guess the simplest functions which have
 vanishing Poisson brackets with $C$:
\begin{equation}
Q_{1}P_{1},\; -Q_{2}P_{2},\; Q_{1}P_{2},\; Q_{2}P_{1},
                                          \label{1.}
\end{equation}
\begin{equation}
-\frac{Q_1}{Q_2},\quad \frac{P_2}{P_{1}}.  \label{2.}
\end{equation}
The regular quadratic expressions (\ref{1.})
will be seen to be related to linear continuous
 symmetries, whereas the singular functions (\ref{2.})
 are connected with the angular variable $\psi$.

Since the angular variable $\psi$ on $\tilde \Gamma$ is
 constant along $C$-orbits, it is another natural
candidate for a symmetry
generator. As an angular variable, it is however singular
at its `axis', so it is not a well-defined function on
the whole of $\tilde \Gamma$. We can still find
 some extensions of $\psi$ to the outside of $\tilde \Gamma$.
Such extensions are not unique;
indeed, using the relations (\ref{14.}) -- (\ref{15.}),
we can immediately define the following two:
\begin{enumerate}
    \item A function  $\psi_{1}$, independent of
 $Q_{1}$ and $Q_{2}$, defined by
\begin{eqnarray*}
\psi_{1}(P_{1},P_{2})& = &  2 \arctan \frac{P_{2}}{P_{1}}
 \quad \mbox{for}\; P_{1} \geq 0, \\
\psi_{1}(P_{1},P_{2}) & = & 2\pi + 2 \arctan
 \frac{P_{2}}{P_{1}} \quad \mbox{for}\; P_{1} \leq 0;
\end{eqnarray*}
the `axis$_{1}$' of  $\psi_{1}$ is the $Q$-plane
($P_{1}=P_{2}=0$) and the corresponding cut$_{1}$
is given by $P_{1}=0$, $P_{2} \leq 0$.
       \item A function $\psi_{2}$, independent of $P_{1}$
and $P_{2}$, defined by
\begin{eqnarray*}
\psi_{2}(Q_{1},Q_{2}) &= & - 2 \arctan \frac{Q_{1}}{Q_{2}}
 \quad \mbox{for}\; Q_{2} \geq 0, \\
\psi_{2}(Q_{1},Q_{2})& = & 2\pi - 2 \arctan
\frac{Q_{1}}{Q_{2}} \quad \mbox{for}\; Q_{2} \leq 0,
\end{eqnarray*}
the `axis$_{2}$' of  $\psi_{2}$ is the $P$-plane
($Q_{1}=Q_{2}=0$) and  $\psi_{2}$  is well-defined only on
$\Gamma \; \backslash \mbox{cut}_{2}$, where cut$_{2}$
is defined by $Q_{2}=0$, $Q_{1} \geq 0$;
limits of $\psi_{2}$ towards the cut$_{2}$ from
($Q_{2}>0$)-side give  $\psi_{2}=- \pi$, those from
($Q_{2}<0$)-side give  $\psi_{2}=3\pi$.
\end{enumerate}

Further, for any smooth periodic function $f$ with the period
$4\pi$, $f(\psi_{\lambda})$, $\lambda$= 1, 2, is a smooth
function on $\Gamma \; \backslash$
 `axis$_{\lambda}$'. The Hamiltonian vector
 field $\xi^{\lambda}_{f}$ of
$f(\psi_{\lambda})$ is given by
$
\xi^{\lambda}_{f} =f'(\psi_{\lambda})\xi^{\lambda},
$
where  $\xi^{\lambda}$ is the Hamiltonian vector
field of  $\psi_{\lambda}$; $\xi^{\lambda}$ is
well-defined on $\Gamma \; \backslash$  `axis$_{\lambda}$',
 as $d\psi_{\lambda}$ is ($d\psi_{\lambda}$
is a closed but non-exact 1-form).

We easily calculate $\xi^\lambda$ which results in
\bdm
\xi^1 = - \frac{2P_2}{P^2_1 +P^2_2} \frac{\partial}{\partial Q_1} +
\frac{2P_1}{P^2_1 + P^2_2} \frac{\partial}{\partial Q_2}
\edm
and
\bdm
\xi^2 = \frac{2Q_2}{Q^2_1 + Q^2_2}\frac{\partial}{\partial P_1} -
\frac{2Q_1}{Q^2_1 + Q^2_2} \frac{\partial}{\partial P_2}.
\edm
Thus, at $\tilde{\Gamma}$, $\xi^1$ and $\xi^2$ are tangent to
$\tilde{\Gamma}$,
$\xi^\lambda$ is singular at axis$_\lambda$ for $\lambda = 1,2$, but it is
complete in $\Gamma \backslash `\mbox{axis}_\lambda$'!

\section{Linear symmetries}
In this section, we will select a particular group of symmetries that is
admitted by our system; it consists of {\em linear} transformations in the
phase space $\Gamma$.

{}From the algebraic point of view, all linear
canonical transformations of $\Gamma$ are given by
the linear action of the symplectic group
$Sp(4,\reell)$ on ${\reell}^4$. Of these, only those are
symmetries which preserve the quadratic
constraint $C=0$ with signature $(+ - - +)$.

In order to find {\it all linear symmetries} we first look
at the intersection of $4 \times 4$ matrix groups
$Sp(4,\reell)$ and $O(2,2)$ acting on $\reell^4$. In the coordinates
$ {\bf X} = (x_n)= (Q_1,Q_2,P_1,P_2)$,
 the conditions on a $4 \times 4$ matrix
${\bf \Lambda}$ to be in $Sp(4,\reell)$ as well as in $O(2,2)$ are
\begin{equation}
{\bf \Lambda}^{T} {\bf \Omega}{\bf \Lambda} = {\bf \Omega}\;
 \quad \mbox{and} \quad
{\bf \Lambda}^{T}{\bf G}{\bf \Lambda} = {\bf G},
                                                 \label{7.}
\end{equation}
where ${\bf \Lambda}^{T}$ is the transposed
matrix, and
\bdm
{\bf \Omega} = \left( \begin{array}{rr} 0 & I \\
              -I & 0 \end{array} \right), \quad
{\bf G}= \left( \begin{array}{rr} 0 & I \\
                I & 0 \end{array} \right),
\edm
with $I$ being the $2 \times 2$ unit matrix.
Writing ${\bf \Lambda}$ in the block form, one can solve
 eqs.\ (\ref{7.}) with the result
\begin{equation}
{\bf \Lambda} =
 \left( \begin{array}{rr} \Lambda & 0 \\
  0 & {\Lambda^{-1}}^{T}\end{array} \right), \label{10.}
\end{equation}
for all $\Lambda \in GL(2,\reell)$.

Since, because of (\ref{0.}), one should assume the
 preservation of $C$ only up to a  factor $\gamma \neq 0$,
eqs.\ (\ref{7.}) can be generalized into equations
\bdm
{\bf \Lambda}^{T} {\bf \Omega}{\bf \Lambda} = {\bf \Omega}\;
\quad \mbox{and} \quad
{\bf \Lambda}^{T}{\bf G}{\bf \Lambda} =\gamma {\bf G}.
\edm
It turns out, however, that these have solutions
only for $\gamma = 1$ and $\gamma = -1$. In the latter case
\bdm
{\bf \Lambda} = \left( \begin{array}{rr} 0 & I \\
              -I & 0 \end{array} \right)
 \left( \begin{array}{rr} \Lambda & 0 \\
         0 & {\Lambda^{-1}}^{T}\end{array} \right),
\quad \forall \; \Lambda \in GL(2,\reell).
\edm
These results suggest that we investigate the continuous
linear symmetries $\Lambda \in GL^+(2,\reell)$ first, and then
turn to an extension by discrete linear symmetries.

\subsection{Continuous linear symmetries}\label{17.}
Writing the group element in the form
 ${\bf \Lambda} = \exp {\bf A}$,
 the conditions for ${\bf A}$ to belong to the
intersection of the corresponding Lie algebras
$sp(4,\reell)$ and $o(2,2)$ become
\bdm
{\bf A}^{T} {\bf \Omega}+ {\bf \Omega}{\bf A} = {\bf 0},
 \qquad  {\bf A}^{T}{\bf G}+ {\bf G}{\bf A} = {\bf 0}.
\edm
By straightforward calculation we obtain the general
element ${\bf A}$ of $sp(4,R) \cap o(2,2)$
of the form
\bdm
{\bf A} = \left( \begin{array}{rr} A & 0 \\
              0 & -A^{T}\end{array} \right),
\edm
where $A$ is an arbitrary real $2 \times 2$ matrix.
Thus ${\bf A}$ can be expanded into
 the real linear combination
\bdm
{\bf A} = c_{0}\ {\bf L}_{0} + c_{1}\ {\bf L}_{1}
         +c_{2}\ {\bf L}_{2} + c\ {\bf Z}
\edm
of 4 particular tensor products of $I$ and
the Pauli matrices $\sigma^{a}$, $a$ = 1, 2, 3:
\begin{eqnarray}
{\bf L}_{0} & = &
    -\frac{i}{2} \sigma^{2} \otimes I, \label{3.}\\
{\bf L}_{1} & = &
       \frac{1}{2} \sigma^{1} \otimes \sigma^{3},  \\
{\bf L}_{2} & = &
\frac{1}{2} \sigma^{3} \otimes \sigma^{3}, \label{4.}\\
{\bf Z}& = &I \otimes \sigma^{3}. \nn
\end{eqnarray}
These matrices satisfy the commutation relations
\bdm
[{\bf L}_{0},{\bf L}_{1}] = -{\bf L}_{2}, \;
[{\bf L}_{0},{\bf L}_{2}] = {\bf L}_{1}, \;
[{\bf L}_{1},{\bf L}_{2}] = {\bf L}_{0},\;
[{\bf Z},{\bf L}_{\kappa}] = {\bf 0}
\edm
of $gl(2,\reell)$.

The generators of the corresponding linear canonical
transformations are bilinear functions of the form
\begin{equation}
f^{{\bf A}} = - \frac{1}{2} \sum_{mn} x_{m}
({\bf \Omega A})_{mn} x_{n}.                  \label{5.}
\end{equation}
Since
\begin{equation}
\{f^{\bf A},f^{\bf B}\}= f^{[{\bf A},{\bf B}]}, \label{6.}
\end{equation}
we have a {\it canonical realization}
of the Lie algebra $gl(2,\reell)$
based on two pairs of canonical variables.

Let us next find a connection of the functions (16) with the perennials
listed in the previous section. We note that the four functions (\ref{1.}) are
all $C^{\infty}$ over $\Gamma$ and the first two of them satisfy
\bdm
Q_{1}P_{1} = - Q_{2}P_{2} + C,
\edm
hence they are not independent functions on
$\tilde{\Gamma}$.
But (\ref{5.}) with (\ref{3.}) -- (\ref{4.}) give the
three independent linear combinations
$f^{{\bf L}_\kappa}$, $\kappa$ = 0, 1, 2, which we shall
denote $L_\kappa$; as generators of extended point
transformations, they can be written in the simple form
\begin{eqnarray}
L_{0}& = &f^{{\bf L}_0} =\frac{1}{2}(Q_{1}P_{2}-Q_{2}P_{1})
       =  Q^{T} \frac{1}{2} i\sigma^{2}P,
                                          \label{8.} \\
L_{1}& =& f^{{\bf L}_1}=\frac{1}{2}(Q_{1}P_{2}+Q_{2}P_{1})
       =  Q^{T} \frac{1}{2}\sigma^{1}P,  \\
L_{2}& =& f^{{\bf L}_2} =\frac{1}{2}(Q_{1}P_{1}-Q_{2}P_{2})
       =  Q^{T} \frac{1}{2}\sigma^{3}P. \label{9.}
\end{eqnarray}
The central element ${\bf Z}$ of the symmetry algebra
corresponds to the constraint function
\bdm
C = f^{\bf Z} = Q_{1}P_{1} + Q_{2}P_{2}=Q^{T}IP.
\edm
In accordance with (\ref{6.}),
the Poisson bracket algebra of the obtained perennials is
\begin{equation}
\{ L_0, L_1\} = - L_2, \;
\{ L_0, L_2\} =  L_1, \;
\{ L_1, L_2\} =  L_0, \;
\{ C ,  L_\kappa\} = 0.      \label{29.}
\end{equation}
 The formulae (\ref{8.}) -- (\ref{9.}) suggest the
use of a simple (anti-)isomorphic representation of
 $sl(2,\reell)$ by real $2 \times 2$ matrices
$\frac{1}{2}\tau_\kappa$, $\kappa =0,1,2$, where
\bdm
\tau_0 = i\sigma^2
=  \left( \begin{array}{rr} 0 & 1 \\
                -1 & 0 \end{array} \right), \;
\tau_1 = \sigma^1
=  \left( \begin{array}{rr} 0 & 1 \\
                1 & 0 \end{array} \right),  \;
\tau_2 = \sigma^3
=  \left( \begin{array}{rr} 1 & 0 \\
                0 & -1 \end{array} \right),
\edm
and
\bdm
f^{\bf L_\kappa} = Q^{T} \frac{1}{2}\tau_{\kappa}P.
\edm
We shall not consider $f^{\bf Z}$ (represented by
 $I$) any more, since it generates the orbits.
 Thus, we restrict $gl(2,\reell)$ to
 $gl(2,\reell)/\reell = sl(2,\reell)$.

At this point we arrived at the Lie algebra of
 the group $SL(2,\reell)$ and of its covering groups.
 In order to decide which of these groups is relevant as
the symmetry of our model, we shall look at the
one-parameter subgroups ${\bf \Lambda}_{\kappa}(t)$ generated by
${\bf L}_{\kappa}$, $\kappa=0, 1, 2$, and their action in
$\Gamma$ and in the space of orbits in $\tilde{\Gamma}$.
By exponentiation we find in the $2\times 2$
representation
\begin{eqnarray*}
\Lambda_{0}(t)& = &\exp (-\frac{i}{2}\sigma^{2} t) =
  I \cos \frac{t}{2} - i\sigma^{2} \sin \frac{t}{2},\\
\Lambda_{1}(t)& = &\exp (\frac{\sigma^1}{2}t) =
  I \cosh \frac{t}{2} + \sigma^{1} \sinh \frac{t}{2},\\
\Lambda_{2}(t)& = &\exp (\frac{\sigma^3}{2}t) =
  I \cosh \frac{t}{2} + \sigma^{3} \sinh \frac{t}{2}.
\end{eqnarray*}

In the phase space $\Gamma$, the actions of the
one-parameter subgroups on the column
vectors ${\bf X}$  with components $(Q_1,Q_2,P_1,P_2)$
are given by $4\times 4$ matrices
 ${\bf \Lambda}_{\kappa}(t) = \exp(t{\bf L}_\kappa)$:
\begin{eqnarray*}
{\bf \Lambda}_{0}(t)&=&{\bf 1} \cos \frac{t}{2} +
     \mbox{diag}(-i\sigma^2,-i\sigma^2)\sin \frac{t}{2}, \\
{\bf \Lambda}_{1}(t)&=&{\bf 1} \cosh \frac{t}{2} +
     \mbox{diag}(\sigma^1,-\sigma^1)\sinh \frac{t}{2}, \\
{\bf \Lambda}_{2}(t)&=&{\bf 1} \cosh \frac{t}{2} +
     \mbox{diag}(\sigma^3,-\sigma^3)\sinh \frac{t}{2}.
\end{eqnarray*}
The integral curves of the actions generated by ${\bf L}_\kappa$
through a point ${\bf X}(0)$ in $\Gamma$ are then given
by
\bdm
{\bf X}(t) = {\bf \Lambda}_{\kappa}(t){\bf X}(0).
\edm
Later also the following relation will be useful:
\bdm
{\bf \Lambda}_{\theta}(t) ={\bf \Lambda}_{0}(-\theta)
   {\bf \Lambda}_{1}(t){\bf \Lambda}_{0}(\theta)=
\exp[({\bf L}_{1}\cos \theta + {\bf L}_{2}\sin \theta)t].
\edm
Since
\bdm
{\bf \Lambda}_{0}(2\pi)= -{\bf 1},\qquad
{\bf \Lambda}_{0}(4\pi)= {\bf \Lambda}_{0}(0)=
{\bf 1},
\edm
we see that the linear action of $SL(2,\reell)$ in $\Gamma$
is, in fact, a {\it faithful representation of}
$SL(2,\reell)$ -- {\it the double covering $SO_2(2,1)$ of}
  $SO(2,1)$.

We have still to describe the action of the group
$SL(2,\reell)$ on $\tilde \Gamma$ and on the orbits.
We recall here that the perennials $L_\kappa$
 are given by
\bdm
L_0 = -\frac{1}{2}xy, \quad
L_1 = \frac{1}{2}xy \cos \psi, \quad
L_2 = -\frac{1}{2}xy \sin \psi.
\edm
The real matrix
(\ref{10.}) corresponds to the
$SL(2,\reell)$-transformation
\bdm
Q' = \Lambda Q, \quad
 P' = {\Lambda^{-1}}^{T}P,
\edm
where we put
\bdm
\Lambda = \left( \begin{array}{rr} \alpha & \beta \\
                \gamma & \delta \end{array} \right),
\quad \alpha \delta - \beta \gamma = 1.
\edm
Substituting this into the definitions (\ref{11.}) --
(\ref{12.}) of $x$, $y$ and $\psi$, one finds that
\bdm
y'  m(\psi') = y  u, \quad
x' n(\psi') = x v,
\edm
where
\begin{eqnarray*}
u & = & \Lambda m(\psi) =
(-\alpha \sin \frac{\psi}{2} + \beta \cos \frac{\psi}{2},
-\gamma \sin \frac{\psi}{2} + \delta \cos \frac{\psi}{2}),
\\ v &= &{\Lambda^{-1}}^{T}  n(\psi) =
(-\gamma \sin \frac{\psi}{2} + \delta \cos \frac{\psi}{2},
\alpha \sin \frac{\psi}{2} - \beta \cos \frac{\psi}{2}).
\end{eqnarray*}
The norms of the vectors $u$ and  $v$
are equal. Denoting them by
\begin{equation}
N(\psi, \Lambda) =
 \sqrt{(-\alpha \sin \frac{\psi}{2} + \beta \cos \frac{\psi}{2})^{2}
+(-\gamma \sin \frac{\psi}{2} + \delta \cos \frac{\psi}{2})^{2} },
                      \label{21.}
\end{equation}
we obtain
\begin{equation}
y' = y N(\psi,\Lambda), \quad x' = x N(\psi,\Lambda).
                                         \label{16.}
\end{equation}
The angle $\psi'$ is a function of $\psi$ and $\Lambda$
only, namely
\begin{equation}
   \psi' =F(\psi,\Lambda) = 2 \arctan
\frac{\alpha \sin \frac{\psi}{2} - \beta \cos \frac{\psi}{2}}
     {-\gamma \sin \frac{\psi}{2} + \delta \cos
     \frac{\psi}{2}}.    \label{22.}
\end{equation}
Thus, the \em transformations of $\psi$ decouple
 from those of $x$ and $y$. \em We observe that
\begin{equation}
\frac{\partial F}{\partial \psi}(\psi, \Lambda) =
\frac{1}{N^{2}(\psi,\Lambda)}.   \label{24.}
\end{equation}

The functions $F(\psi,\Lambda)$ and $N(\psi,\Lambda)$
 satisfy certain composition laws. Namely, if
\bdm
y'' m(\psi'') =
 y'\Lambda' m(\psi'), \quad
 x'' n(\psi'') = x'  \Lambda' n(\psi'),
\edm
the substitution for $x'$, $y'$ and $\psi'$ yields
\begin{eqnarray*}
y'' & = & y N(\psi, \Lambda) N(F(\psi,\Lambda),\Lambda'),\\
x''& = & x N(\psi, \Lambda) N(F(\psi,\Lambda),\Lambda'),\\
\psi'' & = & F(F(\psi,\Lambda),\Lambda').
\end{eqnarray*}
On the other hand, one has
\begin{eqnarray*}
y'' & = & y N(\psi, \Lambda' \Lambda),\\
x'' & = & x N(\psi, \Lambda' \Lambda),\\
\psi'' & = & F(\psi,\Lambda' \Lambda).
\end{eqnarray*}
Hence, we obtain that
\begin{eqnarray}
N(\psi,\Lambda' \Lambda) & = & N(\psi,\Lambda)
                 N(F(\psi,\Lambda),\Lambda'),
                        \label{25.}   \\
F(\psi,\Lambda' \Lambda) &=& F(F(\psi,\Lambda),
           \Lambda').      \label{26.}
\end{eqnarray}

In particular, for the subgroup
\bdm
\Lambda_{0}(t) =
 \left( \begin{array}{rr} \cos \frac{t}{2} & -\sin \frac{t}{2}\\
\sin \frac{t}{2} &  \cos \frac{t}{2} \end{array} \right)
\edm
we have
\be
F_{0}(\psi,t) = \psi + t, \quad N_{0}(\psi,t) =1.
\ee
For the subgroup
\be
\Lambda_{\theta}(t) =
 \left( \begin{array}{cc}
 \cosh \frac{t}{2}+\sin \theta \sinh \frac{t}{2},
 & \cos \theta \sinh \frac{t}{2}\\
\cos \theta \sinh \frac{t}{2},
 &  \cosh \frac{t}{2} - \sin \theta \sinh \frac{t}{2}
\end{array} \right),
\ee
we find in this way
\be
F_{\theta}(\psi,t) = \frac{\pi}{2} - \theta  +
   2\arctan [e^{t} \tan(\frac{\psi +\theta}{2} -
    \frac{\pi}{4})]
\ee
for $\psi \in
    [-\frac{\pi}{2}-\theta,\frac{3\pi}{2}-\theta]$,
\be
F_{\theta}(\psi,t) = \frac{5\pi}{2} - \theta  +
   2\arctan [e^{t} \tan(\frac{\psi +\theta}{2} -
    \frac{\pi}{4})]
\ee
for $\psi \in
    [\frac{3\pi}{2}-\theta,\frac{7\pi}{2}-\theta]$, and
\bdm
N_{\theta}(\psi,t)=\sqrt{\cosh t - \sin(\psi +\theta)\sinh t}.
\edm
For $\theta = 0$ we get the function $F_{1}$
corresponding to $\Lambda_{1}(t)$,
\begin{eqnarray*}
F_{1}(\psi,t)& =& \frac{\pi}{2} +
   2\arctan [e^{t} \tan(\frac{\psi}{2} -
    \frac{\pi}{4})] \;\; \mbox{for}\;\; \psi \in
    [-\frac{\pi}{2},\frac{3\pi}{2}],\\
F_{1}(\psi,t)& =& \frac{5\pi}{2}  +
   2\arctan [e^{t} \tan(\frac{\psi}{2} -
    \frac{\pi}{4})] \;\; \mbox{for} \;\;\psi \in
    [\frac{3\pi}{2},\frac{7\pi}{2}].
\end{eqnarray*}
For $\theta = \frac{\pi}{2}$ we get\footnote{The
functions $F_{\kappa}$, $\kappa$=0, 1, 2, satisfy
the relations
\begin{eqnarray*}
F_{i}(\psi,0)& = & \psi,\\
F_{i}( F_{i}(\psi,t),s) &= & F_{i}(\psi,t + s),\\
F_{1}(\psi,t)& =& F_{2}(\psi-\frac{\pi}{2},t)+\frac{\pi}{2}.
\end{eqnarray*}.}
the function $F_{2}$ corresponding to $\Lambda_{2}(t)$,
\begin{eqnarray*}
F_{2}(\psi,t) &=& 2\arctan (e^{t} \tan\frac{\psi}{2}) \; \;
\mbox{for}\;\; \psi \in  [-\pi, \pi],\\
F_{2}(\psi,t) &=&
 2\pi  +   2\arctan (e^{t} \tan\frac{\psi}{2}
    ) \;\; \mbox{for}\;\; \psi \in [\pi, 3\pi].
\end{eqnarray*}
We also have
\begin{equation}
\frac{\partial F_{1}}{\partial t} = -\cos F_1,\quad
\frac{\partial F_{2}}{\partial t} = \sin F_2.   \label{23.}
\end{equation}

The stratification of $\tilde \Gamma \;
$ under
the  $SL(2,\reell)$-action is given by the following theorem.

\noindent {\bf Theorem.} The  $SL(2,\reell)$-orbits in
$\tilde \Gamma$ are sets
 $\Gamma_{\alpha}$ given by
\bdm
x \cos \alpha + y \sin \alpha = 0, \quad  \;
\alpha \in [0,2\pi).
\edm

{\it Proof.} On the one hand, eqs.\ (\ref{16.}) imply
\bdm
x' \cos \alpha + y' \sin \alpha = 0 \quad
\mbox{iff} \quad
x \cos \alpha + y \sin \alpha = 0
\edm
under all $\Lambda$. Thus the orbits of $SL(2,\reell)$
lie within $\Gamma_{\alpha}$. On the other hand, let
$(x,y,\psi)$ and $(x',y',\psi')$ be two arbitrary points
in $\Gamma_{\alpha}$. Then we can construct a transformation
of $SL(2,\reell)$ transforming  $(x,y,\psi)$ into
 $(x',y',\psi')$.
First we observe that $x^{2}+y^{2}>0$, $x'^{2}+y'^{2}>0$,
as the origin in $\tilde \Gamma$ has been excluded. Then
the transformation is
\bdm
\Lambda_{0}(\psi' - \pi)
 \Lambda_{2}(\log \frac{x'^{2}+y'^{2}}{x^{2}+y^{2}})
\Lambda_{0}(\pi - \psi).
\edm
Indeed, $\Lambda_{0}(\pi - \psi): (x,y,\psi)\mapsto
(x,y,\pi)$. Next, applying $\Lambda_{2}(t)$ at
 $\psi =\pi$, the angle remains unchanged,
 while $x \mapsto x e^{t \over 2}$
and  $y\mapsto y e^{t \over 2}$. The above choice of
$t$ in $\Lambda_{2}$ yields $(x',y',\pi)$. Finally, we apply
$\Lambda_{0}(\psi' - \pi)$ to obtain  $(x',y',\psi')$.
$\Box$

Suppose next that $\gamma_1$ and $\gamma_2$ are
$C$-orbits. Does there exist an
$SL(2,\reell)$-transformation ${\bf \Lambda}$ such that
\bdm
 {\bf \Lambda}\gamma_1  = \gamma_2  ?
\edm
The Theorem implies that ${\bf \Lambda}$ exists
 if and only if
\begin{enumerate}
\item[a)] $L_{0}>0$ along both $\gamma_1$ and
         $\gamma_2 $;
\item[b)]  $L_{0}<0$ along both $\gamma_1$ and
         $\gamma_2 $;
\item[c)] $x=0$ along both $\gamma_1$ and
         $\gamma_2 $;
\item[d)] $y=0$ along both $\gamma_1$ and
         $\gamma_2 $.
\end{enumerate}
Thus, $SL(2,\reell)$ does not act almost transitively
on $\tilde \Gamma$/orbits:
 there are four distinct $SL(2,\reell)$-orbits,
 and two of them [cases a) and b)] are open.

This property is quite general. Suppose that
${\cal G}$ is a {\it connected} group of {\it regular}
symmetries (like $SL(2,\reell)$). Then, because of
 the regularity, no free orbit
can be mapped into an imprisoned orbit.
Suppose that the free orbits form
several disconnected components. Then ${\cal G}$
cannot map $\gamma$ from one component to another one
because of the connectivity.

It is interesting to compare the group theoretic and
algebraic approaches at this point. The perennials
 $ L_0$, $ L_1$ and $ L_2$ separate orbits in the
open subset of $\tilde \Gamma$ defined by $L_{0}
\neq 0$. Thus, from the algebraic point of view, they
form a complete system. However, the group $SL(2,\reell)$
that is generated by them is not transitive!

Finally, the action of the group generated by $f(\psi_\lambda)$ on
$\tilde{\Gamma}$ can be calculated using the form of $\xi^\lambda$ given at
the end of subsection 2.3. We obtain easily
\begin{eqnarray*}
\xi^1 x & =  0, \;\;\; \xi^2 x  =       \frac{2}{y}, \\
\xi^1 y & =  \frac{2}{x},\; \; \; \xi^2 y  =  0,  \\
\xi^1 \psi & =  0, \;\; \;\xi^2 \psi  =  0.
\end{eqnarray*}
Thus $\xi^\lambda$ is complete on $\tilde{\Gamma} \backslash
\mbox{axis}_\lambda$,
$\lambda = 1,2$, $\mbox{axis}_\lambda \cap \tilde{\Gamma}$ being given by $x=
0$ for $\psi_1$ and by $y=0$ for $\psi_2$

\subsection{Discrete linear symmetries}
We have seen that the whole group of linear symmetries
preserving ${\bf G}$ up to a non-vanishing multiplier,
consists of two disconnected sets
\begin{equation}
\{  \left( \begin{array}{rr} \Lambda & 0 \\
        0 & {\Lambda^{-1}}^{T}\end{array} \right)\}
\quad \mbox{and} \quad
\{ \left( \begin{array}{rr} 0 & I \\
              -I & 0 \end{array} \right)
 \left( \begin{array}{rc} \Lambda & 0 \\
      0 & {\Lambda^{-1}}^{T}\end{array} \right)\}
                                    \label{13.}
\end{equation}
with $\Lambda \in GL(2,\reell)$ which itself  consists
of two connected components, $\mbox{sgn}\,\det \Lambda =
 \pm 1$.
Recall that the group $SL(2, \reell)$ of the nontrivial linear symmetries
(which is a subgroup of the first component containing the identity) is not
almost transitive. Let us therefore try to add some discrete transformations to
it.
\begin{itemize}
   \item One choice is the discrete
       transformation from (\ref{13.})
\bdm
\tau = \left( \begin{array}{rr} 0 & I \\
              -I & 0 \end{array} \right)\;
  ( = {\bf \Omega}),
\edm
which reverses the sign of $C$ and allows us to extend
$SL(2,\reell)$ by a disconnected component as in (\ref{13.}).
   \item Another choice is a transformation from
the second component of
$GL(2, R)$ (with negative determinant):
\bdm
\sigma = \mbox{diag}(-\sigma^3, -\sigma^3), \quad
\det \sigma^3 = -1.
\edm
\end{itemize}
The transformations act on the coordinates as follows:
\begin{eqnarray*}
\tau: Q_1 \mapsto + P_1, \; Q_2 \mapsto +P_2, \;
               P_1 \mapsto -Q_1, \; P_2 \mapsto -Q_2, \\
\sigma: Q_1 \mapsto -Q_1, \; Q_2 \mapsto +Q_2, \;
               P_1 \mapsto -P_1, \; P_2 \mapsto +P_2,\\
\end{eqnarray*}
and satisfy the relations
\bdm
\sigma^2 = {\bf 1}, \quad \tau^2=-{\bf 1},\quad
 \tau \sigma= \sigma \tau.
\edm
Evidently, $\tau$ and $\sigma$ generate the Abelian
group $Z_{4} \times Z_{2}$ of order 8.
Also the actions of $\tau$ and $\sigma$  on
$\tilde \Gamma$ can be found easily:
\begin{eqnarray}
\tau&:&x \mapsto y, \quad y \mapsto x, \quad
              \psi \mapsto  \psi - \pi, \;\; \nn  \\
& &(x \mapsto -y, \quad y \mapsto -x, \quad
              \psi \mapsto  \psi + \pi ), \nn \\
\sigma&:&x \mapsto -x, \quad y \mapsto y, \quad
              \psi \mapsto - \psi, \;\; \nn \\
& &(x \mapsto x, \quad y \mapsto -y, \quad
              \psi \mapsto 2\pi - \psi ).\label{18.}
\end{eqnarray}

A straightforward calculation reveals the
automorphisms of $sl(2,\reell)$ induced by
$\tau$ and $\sigma$:
\begin{eqnarray}
  \tau {\bf L}_{0}\tau^{-1} &= & +{\bf L}_{0},\quad
\tau {\bf L}_{1} \tau^{-1} =  -{\bf L}_1,\quad
\tau {\bf L}_{2}\tau^{-1}=  -{\bf L}_{2}, \nn \\
\sigma{\bf L}_{0}\sigma^{-1}& =& - {\bf L}_{0},\quad
\sigma{\bf L}_{1}\sigma^{-1} = - {\bf L}_{1}, \quad
\sigma{\bf L}_{2}\sigma^{-1}=  +{\bf L}_{2}. \label{30.}
\end{eqnarray}
In the adjoint representation of $SL(2,\reell) \approx
SO_{2}(2,1)$, ${\bf \sigma}$ corresponds to the inversion
 of the axis 2.  The {\it extended group} generated by
 ${\bf L}_0$, ${\bf L}_1$, ${\bf L}_2$ and $\sigma$
 can, therefore, be also regarded
as $O^{+}_{2}(2,1)$, the orthochronous Lorentz group in
three dimensions. The group $O^+_2(2,1)$ acts almost transitively on
$\tilde{\Gamma}$/orbit. Let us select this group as our first-class canonical
group.

\section{Transversal surfaces}
We have seen in Sec.\ 2 that there is no global transversal surface. In the
present paper, we will try to use instead a system of maximal transversal
surfaces such that each orbit is intersected by at least one surface of the
system. In this section, we describe an example of such a system.

Let us choose the transversal surfaces
$\Gamma_{1}$  and $\Gamma_{2}$ by
\begin{eqnarray*}
\Gamma_{1}=\{(x,y,\psi)\in {\tilde \Gamma} \vert
x = T, \quad  -\infty <y<\infty, \quad
 0\leq \psi <4\pi \},\\
\Gamma_{2}=\{(x,y,\psi)\in {\tilde \Gamma} \vert
y = T, \quad -\infty <x<\infty, \quad
 0\leq \psi <4\pi\},
\end{eqnarray*}
$T$ being an arbitrary positive constant.
The domain of $\Gamma_{1}$ is the subset
$D(\Gamma_{1})$ of $\tilde \Gamma$
satisfying $x \neq 0$, while the domain of  $\Gamma_{2}$ is
that satisfying $y \neq 0$. Thus,  $\Gamma_{1}$
and $\Gamma_{2}$ together intersect all orbits.
It is also clear that  $\Gamma_{1}$ and
$\Gamma_{2}$ are both maximal.

The topology of  $\Gamma_{1}$ or  $\Gamma_{2}$
is $\reell \times S^{1}$. We can introduce coordinates
$(l_{i}, \varphi_{i})$ on  $\Gamma_{i}$ by the
following embedding formulae. For  $\Gamma_{1}$
in $\tilde \Gamma$,
\bdm
x = T, \quad y = -\frac{2l_1}{T}, \quad
\psi = \varphi_1
\edm
and in $\Gamma$,
\begin{eqnarray*}
Q_{1} & = & \frac{2l_1}{T}\sin\frac{\varphi_1}{2},
\quad
Q_{2}  =  -\frac{2l_1}{T}\cos\frac{\varphi_1}{2},
\\
P_1 & = & T \cos \frac{\varphi_1}{2}, \quad
P_2  =  T \sin \frac{\varphi_1}{2};
\end{eqnarray*}
 for  $\Gamma_{2}$  in $\tilde \Gamma$,
\bdm
 x = -\frac{2l_2}{T}, \quad y = T, \quad
\psi = \varphi_2
\edm
and in $\Gamma$,
\begin{eqnarray*}
Q_1 & =&  -T \sin \frac{\varphi_2}{2}, \quad
Q_2  =  T \cos \frac{\varphi_2}{2},\\
P_{1}  &= & -\frac{2l_2}{T}\cos\frac{\varphi_2}{2},
\quad
P_{2}  =  -\frac{2l_2}{T}\sin\frac{\varphi_2}{2},
\end{eqnarray*}
Then the corresponding pull-backs of the symplectic
form $\Omega = dP_{1}\wedge dQ_{1} +
 dP_{2}\wedge dQ_{2}$ in $\Gamma$ are
\bdm
\Omega_i = dl_{i}\wedge d\varphi_{i}, \quad
i = 1,2.
\edm
Thus, $(\Gamma_i, \Omega_i)$ can be considered
as cotangent bundles, $T^{*}{\cal C}_i$, with
the Liouville forms $\theta_i = l_{i}d\varphi_i$,
so that $\Omega_i = d\theta_i$ and ${\cal C}_i
\approx S^1$.

Observe that the orbits in $D(\Gamma_i)$
can be distinguished by the values of two
perennials, $L_0=-xy/2$ and $\psi$. Their values
on   $\Gamma_i$ are given by
\bdm
L_{0}|_{\Gamma_i} = l_i, \quad
\psi |_{\Gamma_i} = \varphi_i.
\edm

Let us define the subsets $\Gamma^\pm_i, \Gamma^0_i$ of $\Gamma_i$ as follows
\begin{eqnarray*}
\Gamma_{i}^+ &= & \{p \in \Gamma_{i}\vert
          l_{i}(p) > 0 \}, \\
\Gamma_{i}^0 &= & \{p \in \Gamma_{i}\vert
          l_{i}(p) = 0 \}, \\
\Gamma_{i}^- &= & \{p \in \Gamma_{i}\vert
          l_{i}(p) < 0 \}.
\end{eqnarray*}
Each point $(l_1,\varphi_1)$ of $\Gamma_{1}^+$
determines  a unique orbit $\gamma$ with the
values of $L_0 = l_1$ and $\psi = \varphi _1$
 along it. The orbit $\gamma$ intersects the
surface $y = -T$ at the point
\bdm
x = \frac{2l_1}{T}, \quad y = -T, \quad
\psi = \varphi_1,
\edm
because $L_0 = - {1 \over 2}xy$, $x$ is negative
at the intersection $\Gamma_1 \cap \gamma$, and
neither $x$ nor $y$ can change sign along an orbit.
Thus, the intersection $\gamma \cap \Gamma_2$
has the coordinates
\bdm
x = -\frac{2l_1}{T}, \quad y = T, \quad
\psi = \varphi_1 + 2\pi,
\edm
 which corresponds to
\begin{equation}
 l_2 = l_1, \qquad
\varphi_2 = \varphi_1 + 2\pi.   \label{19.}
\end{equation}
The mapping $\rho_+ : \Gamma_{1}^{+} \rightarrow
 \Gamma_{2}^{+}$ defined by eq.\ (\ref{19.})
is a symplectic diffeomorphism. Similarly,
we obtain $\rho_- : \Gamma_{1}^{-} \rightarrow
 \Gamma_{2}^{-}$ which is given by
\bdm
 l_2 = l_1, \qquad
\varphi_2 = \varphi_1.
\edm
We also introduce the combined map $\rho$ of
$\Gamma_{1}^{+} \cup \Gamma_{1}^{-}$ onto
$\Gamma_{2}^{+} \cup \Gamma_{2}^{-}$; it will
be called a {\it pasting map}.

We observe finally that none of the symplectic
diffeomorphisms $\rho_{\pm}$ and $\rho$ can be
considered as a ``lift''\footnote{Equivalently,
 $\rho_{\pm}$ and $\rho$ are canonical transformations
which are not extended point transformations.}
 of a diffeomorphism
between the configuration spaces ${\cal C}_1$
and  ${\cal C}_2$ ($\approx S^1$).
Thus, our pasting is rather different from that
encountered in  gauge theories with non-trivial
fibre bundles, where patches of a configuration
space are pasted together by  gauge transformations
\cite{8,9}. This is due to the fact that
the transformation generated by constraints of
gauge fields which are linear in momenta, leaves
configuration space invariant, whereas our constraint
is quadratic in momenta.

Next we will project the symmetry transformations to the transversal surfaces,
as it is described in I.

Let  $a_{i}(\Lambda)$, $i=1,2$, denote the (non-linear)
 action of a symmetry
${\bf \Lambda} : \Gamma \rightarrow \Gamma$ on
$\Gamma_i$ (for the definitions see I;
 $a_{i}(\Lambda)$ are canonical
 transformations on $(\Gamma_i,\Omega_i)$). For the group
$SL(2,\reell)$, we easily read off from the relations
of Sec.\ \ref{17.} that
\begin{equation}
a_{i}(\Lambda)(l_i,\varphi_i) =
(l_{i}N^{2}(\varphi_i,\Lambda),
F(\varphi_i, \Lambda)).\label{20.}
\end{equation}
Thus the sign of $l_i$ is not changed by any
transformation of $SL(2,\reell)$ and \em the
sets $\Gamma_{i}^{\pm}$ and $\Gamma_{i}^{0}$
are  $a_{i}(SL(2,\reell))$-invariant \em.
Moreover, these sets are {\it orbits} of
  $a_{i}(SL(2,\reell))$; this follows easily
from the Theorem of Sec.\ \ref{17.}. In particular,
the action $a_i$ of $SL(2,\reell)$ is
transitive in each of the (overlapping) sets
 $\Gamma_{i}^{\pm}$.

Another important observation is that both sets
 $\Gamma_{1}^{+}$ and  $\Gamma_{2}^{+}$
which are `pasted together' by $\rho_+$ have
the property that $L_{0}\vert_{\Gamma_{i}^{+}}=l_{i}
>0$, $i=1,2$. Similarly,
$L_{0}\vert_{\Gamma_{i}^{-}}=l_{i}<0$.
Now, $L_0$ as a function on $\Gamma_{i}$
generates, via Poisson brackets, the one-parameter
group $a_{i}(\Lambda_{0}(t))$ \cite{1}. Thus,
we will try to identify the relevant representations
$R$ of  $SL(2,\reell)$ by the sign of eigenvalues
of the corresponding generator of
$R(\Lambda_{0}(t))$.

The action of the discrete transformation $\sigma$
on $\tilde\Gamma$ is given by (\ref{18.}). Thus, on
$\Gamma_1$,
\bdm
a_{1}(\sigma)\; : \; l_1 \mapsto -l_1, \;
           \varphi_1 \mapsto 2\pi - \varphi_1,
\edm
and on $\Gamma_2$
\bdm
a_{2}(\sigma)\; : \; l_2 \mapsto -l_2, \;
           \varphi_2 \mapsto -\varphi_2.
\edm
Note that both  $a_{i}(\sigma)$ are extended point
transformations. Moreover, the actions  $a_i$ of
$O_{2}^{+}(2,1)$ (2 connected components)
are {\it almost transitive} on $\Gamma_i$,
the corresponding orbits being
 $\Gamma_{i}^{+}\cup  \Gamma_{i}^{-} $
(2 connected components)
and $\Gamma_{i}^{0}$.

The infinitesimal generators of the transformations
(\ref{20.}) are easily calculated using the relations
(\ref{21.}), (\ref{22.}) and (\ref{23.}):
\begin{eqnarray*}
da_{i}(\frac{\tau_0}{2}) & =&  \frac{\partial}{\partial\varphi_i},\\
da_{i}(\frac{\tau_1}{2}) & = & -\sin\varphi_{i}l_{i}
\frac{\partial}{\partial l_i} -\cos\varphi_{i}
\frac{\partial}{\partial\varphi_i},\\
da_{i}(\frac{\tau_2}{2}) & = & -\cos\varphi_{i}l_{i}
\frac{\partial}{\partial l_i} +\sin\varphi_{i}
\frac{\partial}{\partial\varphi_i};
\end{eqnarray*}
they are global Hamiltonian vector fields of the
following functions on $\Gamma_i$:
\begin{eqnarray*}
L_{0}^{i} & = &\; l_i,\\
L_{1}^{i} & = & -l_{i}\cos\varphi_i,\\
L_{2}^{i} & = & \; l_{i}\sin\varphi_i.
\end{eqnarray*}
The functions $L_{\kappa}^{i}$ are just the restrictions
of the original generators $L_\kappa$ to $\Gamma_i$
(cf. (\ref{8.}) -- (\ref{9.})).

To summarize: we have reduced the original constrained
system to two overlapping unconstrained systems
 ${\cal S}_1$ and ${\cal S}_2$ living on the
 reduced phase spaces  $\Gamma_1$ and $\Gamma_2$,
\bdm
 \Gamma_i \approx T^{*}S^{1} = \{ (l_i,\varphi_i)
 \vert l_i \in \reell, \varphi_i \in [0,4\pi)\},
\edm
with the symplectic forms $\Omega_i = dl_i \wedge
d\varphi_i$. The {\em perennials} of the original system define {\em
observables} for $S_1$ and $S_2$ by projection to the surfaces $\Gamma_1$ and
$\Gamma_2$ as described in I:
 \be
L_{0}^{i}=l_i, \quad L_{1}^{i}=-l_i \cos
\varphi_i, \quad  L_{2}^{i}=l_i \sin
\varphi_i,
\ee
and the parity
\begin{equation}
a_{1}(\sigma) : \; l_1 \mapsto -l_1,\;
      \varphi_1 \mapsto 2\pi - \varphi_1,\quad
a_{2}(\sigma) : \; l_2 \mapsto -l_2,\;
      \varphi_2 \mapsto  - \varphi_2. \label{133}
\end{equation}
The perennial $f(\psi_\lambda)\; (\lambda = 1,2)$, where $f$ is any $4
\pi$-periodic function of its argument, can be projected only to the
transaversal surface $\Gamma_\lambda\;\; (i= \lambda)$.
The projection is given by
\bdm
f(\psi_\lambda) \vert_{\Gamma_\lambda} = f(\psi_\lambda), \hspace{0.3cm} (i =
\lambda) \hspace{0.3cm} \lambda = 1,2.
\edm
We stress that there is no differentiable projection of $f(\psi_1)$ to
$\Gamma_2$ or of $f(\psi_2)$ to $\Gamma_1$!

 There is a relation among the classical observables, $L^i_0$, $L^i_1$
and $L^i_2$:
 \bdm
-(L_{0}^{i})^{2} +(L_{1}^{i})^{2}+(L_{2}^{i})^{2} =0.
\edm
These and $a_i (\sigma)$ generate the group $O_{2}^{+}(2,1)$.

 The overlapping between $S_1$ and $S_2$ is defined by two maps:
\bdm
\rho_+ : \; \Gamma_{1}^+ \rightarrow \Gamma_{2}^+ ,
\quad l_1>0, \; l_2 >0,
\edm
given by the functions
\bdm
l_{2}(l_1,\varphi_1)=l_1, \quad
\varphi_{2}(l_1,\varphi_1)=\varphi_1 +2\pi,
\edm
and
\bdm
\rho_- : \; \Gamma_{1}^- \rightarrow \Gamma_{2}^- ,
\quad l_1<0, \; l_2 <0,
\edm
given by the functions
\bdm
l_{2}(l_1,\varphi_1)=l_1, \quad
\varphi_{2}(l_1,\varphi_1)=\varphi_1.
\edm

\section{Canonical quantization}       \label{40.}
The next step in the group quantization method is to
find, for systems ${\cal S}_i$, suitable unitary
 representations\footnote{The theory of
 representations of the group $SL(2,\reell)$
and its covering groups can be found, e.g.
 in \cite{10}, \cite{11} and \cite{12}.}
of the group $O_{2}^{+}(2,1)$ defining the corresponding
Hilbert spaces ${\cal H}_i$ and quantum observables
as hermitian operators in  ${\cal H}_i$.

 Here we study the representations
of the group that can be found by a straightforward
application of canonical quantization method to
the reduced systems ${\cal S}_i$.
Thus, as our Hilbert spaces we choose the spaces
of square-integrable functions on the configuration
 space $S^1$,
\bdm
\overline{{\cal H}_i}=L^{2}(S^1,d\varphi_i),
\edm
i.e. with the inner product
\begin{equation}
(\Psi,\Phi)_i = \frac{1}{4\pi}\int_{0}^{4\pi}
\overline{\Psi}(\varphi_i)\Phi (\varphi_i)d\varphi_i. \label{140}
\end{equation}
An obvious symmetrization gives the hermitian
operators for the observables $L_{\kappa}^{i}$,
\begin{eqnarray}
{\hat L}_{0}^{i} & = & -i\frac{\partial}{\partial
    \varphi_i},           \label{27.}\\
{\hat L}_{1}^{i} & = & i \cos \varphi_{i}
\frac{\partial}{\partial\varphi_i}-\frac{i}{2}
\sin \varphi_i,  \\
{\hat L}_{2}^{i} & = & -i \sin \varphi_{i}
\frac{\partial}{\partial\varphi_i}-\frac{i}{2}
\cos \varphi_i.              \label{28.}
\end{eqnarray}

We show that the corresponding unitary representation
of $SL(2,\reell)$ is given by
\be
[U(\Lambda)\Psi](\varphi)=\frac{1}{N(\varphi,\Lambda)}
             \Psi(F(\varphi,\Lambda)).
\ee
$U(\Lambda)$ is unitary, since
\begin{eqnarray*}
(U(\Lambda)\Psi,U(\Lambda)\Phi) & = &
 \frac{1}{4\pi}\int_{0}^{4\pi}\frac{1}{N^{2}(\varphi,\Lambda)}
\overline{\Psi}(F(\varphi,\Lambda))\Phi
 (F(\varphi,\Lambda))d\varphi \\
& = &  \frac{1}{4\pi}\int_{F(0,\Lambda)}^{F(4\pi,\Lambda)}
\overline{\Psi}(F)\Phi (F)dF = (\Psi,\Phi)
\end{eqnarray*}
due to (\ref{24.}). For the composition of two group
elements $\Lambda$,  $\Lambda'$ we obtain
\begin{eqnarray*}
[U(\Lambda')(U(\Lambda)\Psi)](\varphi) & = &
U(\Lambda')\frac{1}{N(\varphi,\Lambda)}
\Psi(F(\varphi,\Lambda)) \\
& = & \frac{1}{N(\varphi,\Lambda')}\
\frac{1}{N(F(\varphi,\Lambda'),\Lambda)}
\Psi(F(F(\varphi, \Lambda'),\Lambda)) \\
& = & \frac{1}{N(\varphi,\Lambda\Lambda')}
\Psi(F(\varphi,\Lambda\Lambda')) =
(U(\Lambda\Lambda')\Psi)(\varphi)
\end{eqnarray*}
because of (26) and (27). Finally,
specializing to the subgroups $\Lambda_{0}(t)$
and $\Lambda_{\theta}(t)$ we find that
\begin{eqnarray*}
(U(\Lambda_{0}(t))\Psi )(\varphi ) & = & \Psi (\varphi + t),\\
(U(\Lambda_{\theta}(t))\Psi )(\varphi ) &=&
\frac{1}{N_{\theta}(\varphi +t)} \Psi (F_{\theta}(\varphi +t)).
\end{eqnarray*}
The $t$-derivatives of these functions at $t=0$
are
\begin{eqnarray}
i{\hat L}_{0}\Psi(\varphi) &=& \frac{d}{dt}
[U(\Lambda_{0}(t))\Psi ](\varphi)\vert_{t=0} =
\Psi'(\varphi),\nn \\
i{\hat L}_{\theta}\Psi(\varphi) &=& \frac{d}{dt}
[U(\Lambda_{0}(t))\Psi](\varphi)\vert_{t=0} \nn \\ &=&
-\cos (\varphi +\theta)\Psi'(\varphi) +
\frac{1}{2}\sin (\varphi +\theta) \Psi (\varphi), \label{145}
\end{eqnarray}
which finishes the proof, since the result is
in accordance  with (\ref{27.}) -- (\ref{28.}).
 Observe that $SL(2,\reell)$ acts from the right!

As $a_{i}(\sigma)$ are extended point transformations
whose projections on the configuration spaces
${\cal C}_i$ leave the integration measure invariant,
we can choose for the corresponding operators
${\hat \sigma}_i$
\begin{eqnarray}
({\hat \sigma}_{1}\Psi )(\varphi_1)&=&
\Psi (2\pi - \varphi_1),\quad \varphi_1 \in
   [0,2\pi),                  \label{44.}   \\
&=& \Psi (6\pi - \varphi_1),\quad \varphi_1 \in
[2\pi,4\pi),\\
({\hat \sigma}_{2}\Psi )(\varphi_2) &=&
\Psi(4\pi - \varphi_2),\quad \varphi_2 \in
   [0,4\pi).                    \label{45.}
\end{eqnarray}

It is easy to verify that the commutation relations
correponding to eqs.\ (\ref{29.}) and (\ref{30.})
are satisfied.
Thus, we have hermitian representations of
the Lie algebra and with it unitary representations
of the group $O_{2}^{+}(2,1)$ in
 ${\overline {\cal H}}_i$.

\subsection{Representations of  $O_{2}^{+}(2,1)$}\label{41.}
We want to clarify which of the known representations
of $O_{2}^{+}(2,1)$ we have (cf. \cite{10}, \cite{11} and
\cite{12}), as well as if the representations are
faithful and irreducible.  Taking the
eigenfunctions of ${\hat L}_{0}^{i}$
\bdm
{\hat L}_{0}^{i}\Psi_m = \frac{m}{2}\Psi_m, \quad
\Psi_{m}(\varphi_i) = e^{\frac{i}{2}m\varphi_i},
\quad m \in Z,
\edm
where $Z$ is the set of all integers, as suitable basis in ${\overline {\cal
H}}_i$, we find for the shift operators
\begin{eqnarray}
{\hat L}_{\pm}^{i} = {\hat L}_{1}^{i}\mp
 i{\hat L}_{2}^{i} = i e^{\pm\frac{i\varphi_i}{2}}
 \frac{\partial}{\partial\varphi_i}
 e^{\pm\frac{i\varphi_i}{2}}, \label{42.}  \\
{\hat L}_{\pm}^{i}\Psi_m =
 -\frac{1}{2}(m\pm 1)\Psi_{m\pm 2}.\label{43.}
\end{eqnarray}
Suppose that $m$ is {\it even}. Then
\bdm
{\hat L}_{\pm}^{i}\Psi_m \neq 0
\edm
and all such functions span one irreducible representation.
The representation space ${\cal H}_{i}^{0} \subset
{\overline {\cal H}}_{i}$ contains {\it all $2\pi$-periodic
elements of ${\overline {\cal H}}_{i}$}. Comparison with
\cite{12} reveals that the representation of
$SL(2,\reell)$ realized in  ${\cal H}_{i}^{0}$ is
$C^{0}_{-\frac{1}{4}}$ belonging to the principal series
(with the value of the Casimir operator $q=-\frac{1}{4}$).

Suppose now that $m$ is {\it odd}. Then
\begin{eqnarray*}
{\hat L}_{+}^{i}\Psi_m &=& 0 \quad for \; m = -1,\\
{\hat L}_{-}^{i}\Psi_m &=& 0 \quad for \; m = +1.
\end{eqnarray*}
Hence we obtain two irreducible representations of
$SL(2,\reell)$, one spanned by the functions $\Psi_m$
for $m=-1,-2,\ldots$, and the other for
 $m=1,2,\ldots$. Let us denote the corresponding
subspaces of ${\overline {\cal H}}_{i}$ by
 ${\cal H}_{i}^{+}$ and ${\cal H}_{i}^{-}$,
respectively. The corresponding representations
$R_{i}^{+} = D_{\frac{1}{2}}^{-}$ and
 $R_{i}^{-} =   D_{\frac{1}{2}}^{+}$
belong to the discrete series (the value of the
Casimir operator is again $q=-\frac{1}{4}$ \cite{12}.)

As a result we obtained the orthogonal direct sum
 decomposition
\bdm
{\overline {\cal H}}_{i}={\cal H}_{i}^{0}\oplus
{\cal H}_{i}^{+}\oplus {\cal H}_{i}^{-}
\edm
and observe that, of the three irreducible representations
of $SL(2,\reell)$, only $C^{0}_{-\frac{1}{4}}$ in
 ${\cal H}_{i}^{0}$ is {\it not faithful}, as the
rotation by $2\pi$, $\varphi_i \mapsto \varphi_{i}
+ 2\pi$, is represented by the identity. Thus we have
to throw the even $m$ subspace  ${\cal H}_{i}^{0}$
away.

Consider now the action of the parity operators
${\hat \sigma}_i$ on the basis functions $\Psi_m$
with $m$ odd. We obtain
\be
{\hat \sigma}_{1}\Psi_{m} = - \Psi_{-m}, \quad
{\hat \sigma}_{2}\Psi_{m} =  \Psi_{-m}.\label{46.}
\ee
Thus the operators $\hat{\sigma}_i$ map
${\cal H}_{i}^{+}$ onto ${\cal H}_{i}^{-}$ and
vice versa, so the group $O^{+}_{2}(2,1)$
is represented irreducibly and faithfully on
\bdm
{\cal H}_{i}:={\cal H}_{i}^{+}
               \oplus {\cal H}_{i}^{-}.
\edm
The elements of ${\cal H}_{i}$ are all
{\it $2\pi$-antiperiodic} functions of $L^{2}(S^1)$.

We have found the two irreducible
representations by canonical
quantization; let us close this section by looking
to see whether or not they satisfy the principles of
algebraic and group quantization.
\begin{enumerate}
\item Are all subalgebras of elementary observables
represented irreducibly?\\ This is not the case, as
${\hat L}_{\kappa}^{i}$, $\kappa = 0,1,2$, form such
a subalgebra, and ${\cal H}_{i}$ is reducible.
Observe that any of the three Hilbert spaces
${\cal H}_{i}^{0}$, ${\cal H}_{i}^{+}$ and
${\cal H}_{i}^{-}$ is a possible choice for the
algebraic quantization.
\item Are all relations satisfied?\\
Not quite! If we apply the prescription given in I (cf. also \cite{3}) we
obtain
that $-( \hat{L}^i_0)^2 + (\hat{L}^i_1)^2 + (\hat{L}^i_2)^2 = \frac{1}{4}$.
 \item Are all canonical subgroups represented
irreducibly?\\
They are! Indeed, we have the (minimal) canonical
group $O^{+}_{2}(2,1)$ and in the spaces
${\cal H}_{i}$ an irreducible representation is
realized.
\end{enumerate}

\section{Pasting the Hilbert spaces}

In the previous section, we have obtained {\em two} quantum theories: one for
each of the reduced systems $S_1$ and $S_2$ that correspond to the two maximal
transversal surfaces $\Gamma_1$ and $\Gamma_2$. In the present section, we try
to construct a single quantum theory from the two.

Let us compare the action of the group $SL(2,\reell)$
in the classical phase spaces $\Gamma_i$, with the
representations of  $SL(2,\reell)$ in the quantum
Hilbert spaces ${\cal H}_{1}$ and ${\cal H}_{2}$.
First, the phase spaces have submanifolds
$\Gamma_{i}^+$ and $\Gamma_{i}^-$ such that
$SL(2,\reell)$ acts transitively inside each of them.
Similarly,  ${\cal H}_{i}$ has two subspaces
 ${\cal H}_{i}^{\pm}$ in which the representations
$R_{i}^\pm$ of  $SL(2,\reell)$ are irreducible.
Second, there are pasting maps,
$\rho_+ : \Gamma_{1}^+ \rightarrow \Gamma_{2}^+$
and $\rho_- : \Gamma_{1}^- \rightarrow \Gamma_{2}^-$
which are {\it symplectic} and equivariant with the action
of  $SL(2,\reell)$. That is, let $a_{i}^{\pm}(\Lambda):
\Gamma_{i}^\pm \rightarrow \Gamma_{i}^\pm$,
 $\Lambda \in SL(2,\reell)$, be the corresponding
 maps in  $\Gamma_{i}^\pm$; then
\bdm
a_{2}^{\pm}(\Lambda)=\rho_{\pm}\circ
 a_{1}^{\pm}(\Lambda) \circ \rho_{\pm}^{-1}.
\edm

We can ask for quantum pasting maps,
$U_{\pm}: {\cal H}_{1}^{\pm}
\rightarrow {\cal H}_{2}^{\pm}$ having analogous
properties: $U_{\pm}$ should be {\it unitary}
intertwiners such that
\begin{equation}
R^{\pm}_{2}=U_{\pm}\circ R^{\pm}_{1}
  \circ U_{\pm}^{-1}.            \label{33.}
\end{equation}
Such a relation presupposes that  $R^{\pm}_{1}$ is
 unitarily equivalent to  $R^{\pm}_{2}$ .
 If this is the case,
then the maps  $U_{\pm}$ are determined uniquely up
to phase factors (because of Schur's lemma):
\bdm
U_{\pm}\Psi^{\pm}(\varphi_1) =_{F}
 e^{i\lambda_{\pm}}\Psi^{\pm}(\varphi_2),\label{31.}
\edm
where $\lambda_{\pm}$ are arbitrary real numbers from the
interval $[0,2\pi)$, $\Psi^{\pm}(\varphi_1)$
is an arbitrary element of ${\cal H}_{1}^{\pm}$
and the equality `$=_{F}$' is to be understood as
follows: the symbols $\Psi^\pm$ on both sides denote
the same mapping $S^1 \rightarrow C$.

We are going to paste the Hilbert spaces
 ${\cal H}_{1}$ and ${\cal H}_{2}$ by means of
 $U_{\pm}$ in analogy to pasting the symplectic
manifolds $\Gamma_1$ and $\Gamma_2$ by means of
$\rho_{\pm}$. Let us first give a general definition
and properties.

\noindent {\bf Definition}:
Let  ${\cal H}_{1}$ and  ${\cal H}_{2}$ be two
Hilbert spaces with orthogonal direct sum
decompositions
\begin{eqnarray*}
{\cal H}_{1}&=&{\cal H}_{1}^{0}\oplus
        {\cal H}_{1}',\\
{\cal H}_{2}&=&{\cal H}_{2}^{0}\oplus
        {\cal H}_{2}'.
\end{eqnarray*}
Let $U:{\cal H}_{1}^{0}\rightarrow {\cal H}_{2}^{0}$
be a unitary map. Let ${\cal H}$ be a Hilbert space
defined by
\bdm
{\cal H}={\cal H}_{1}^{0}\oplus
        {\cal H}_{1}'\oplus  {\cal H}_{2}',
\edm
and let two unitary maps $U_1:{\cal H}_{1}
\rightarrow {\cal H}$, $U_2:{\cal H}_{2}
\rightarrow {\cal H}$ be defined by
\begin{eqnarray*}
U_{1}(\Psi_{1}^{0};\Psi_{1}')&=&(\Psi_{1}^{0};
\Psi_{1}';0), \\
U_{2}(\Psi_{2}^{0};\Psi_{2}')&=&
 (U^{-1}\Psi_{2}^{0};0;\Psi_{2}'),
\end{eqnarray*}
where $\Psi_{1}^{0}\in {\cal H}_{1}^{0}$,
$\Psi_{2}^{0}\in {\cal H}_{2}^{0}$,
$\Psi_{1}'\in {\cal H}_{1}'$,
$\Psi_{2}'\in {\cal H}_{2}'$,
and (.;.;.)  denotes an element of the
orthogonal direct sum. Then the set
$( {\cal H}_{1},{\cal H}_{2},U_1,U_2,{\cal H})$
is called {\it pasting of ${\cal H}_{1}$ and ${\cal H}_{2}$
by means of $U$}.

\noindent {\bf Lemma}: Let $V_1: {\cal H}_{1}^{0}
\rightarrow {\cal H}_{1}^{0}$ and $V_2:{\cal H}_{2}^{0}
\rightarrow {\cal H}_{2}^{0}$ be two maps satisfying
the condition
\begin{equation}
V_{2} \circ U = U \circ V_{1}.       \label{32.}
\end{equation}
Then
\bdm
U_{1} \circ V_{1} \circ U_{1}^{-1} =
U_{2} \circ V_{2} \circ U_{2}^{-1}.
\edm

\noindent {\it Proof}. As the domain of $V_1$ is
${\cal H}_{1}^{0}$, the domain of
$U_{1} \circ V_{1} \circ U_{1}^{-1}$ is
\bdm
U_{1}{\cal H}_{1}^{0}=\{(\Psi_{1}^{0};0;0)\vert
\Psi_{1}^{0} \in {\cal H}_{1}^{0}\},
\edm
and we have
\bdm
(U_{1} \circ V_{1} \circ U_{1}^{-1})
(\Psi_{1}^{0};0;0)=(V_{1}\Psi_{1}^{0};0;0).
\edm
Similarly, the domain of $U_{2} \circ V_{2} \circ
 U_{2}^{-1}$ is
\bdm
U_{2}{\cal H}_{2}^{0}=\{(U^{-1}\Psi_{2}^{0};0;0)\vert
\Psi_{2}^{0} \in {\cal H}_{2}^{0}\},
\edm
and
\bdm
(U_{2} \circ V_{2} \circ U_{2}^{-1})
(U^{-1}\Psi_{2}^{0};0;0)=(U^{-1}V_{2}\Psi_{2}^{0};0;0).
\edm
Thus, if $U^{-1}\Psi_{2}^{0}=\Psi_{1}^{0}$, then
\bdm
(U_{2} \circ V_{2} \circ U_{2}^{-1})
(U^{-1}\Psi_{2}^{0};0;0)=(U^{-1}V_{2}U\Psi_{1}^{0};0;0)
=(V_{1}\Psi_{1}^{0};0;0).
\edm
Q.E.D.

Returning to our system, we can set
 \begin{eqnarray}
{\cal H}_{1}^{0}&=&{\cal H}_{1}={\cal H}_{1}^{+}
\oplus {\cal H}_{1}^{-},\nn \\
{\cal H}_{2}^{0}&=&{\cal H}_{2}={\cal H}_{2}^{+}
\oplus {\cal H}_{2}^{-},\nn \\
U(\Psi_{1}^{+};\Psi_{1}^{-})&=&
(U_{+}\Psi_{1}^{+};U_{-}\Psi_{1}^{-}),  \label{36.}\\
{\cal H}^{+}&=&{\cal H}_{1}^{+},\;\;
{\cal H}^{-}={\cal H}_{1}^{-},\nn \\
{\cal H}&=&{\cal H}^{+}\oplus {\cal H}^{-},\nn \\
U_{1} &=& \mbox{id},\;\; U_2=U^{-1}.      \label{37.}
\end{eqnarray}
Observe that this is a very special case
of the definition above.
For $V_1 = R_{1}(\Lambda)$, $V_2 = R_{2}(\Lambda)$,
where $\Lambda \in SL(2,\reell)$, the condition
(\ref{32.}) is satisfied for any $\lambda_+$ and
$\lambda_-$, because of (\ref{33.}).

Consider next ${\hat \sigma}_1$ and ${\hat \sigma}_2$.
Let $\{\Psi_{m}(\varphi_i)\vert m \in Z_+\}$ be a basis
of ${\cal H}_{i}^{+}$, and
 $\{\Psi_{m}(\varphi_i)\vert m \in Z_-\}$ be a basis
of  ${\cal H}_{i}^{-}$. We obtain then from (\ref{31.})
that
\begin{equation}
U_{\pm}\Psi_{\pm m}(\varphi_1) =_{F}
 e^{i\lambda_{\pm}}\Psi_{\pm m}(\varphi_2) \quad
\forall \; m>0.                              \label{35.}
\end{equation}
Hence, for $m>0$, and $\Psi_{m}(\varphi_1) \in
{\cal H}_{1}^{+}$:
\begin{eqnarray*}
U({\hat \sigma_1}(\Psi_{m}(\varphi_1))) &=&
U(-\Psi_{-m}(\varphi_1)) =_{F} -e^{i\lambda_{-}}
\Psi_{-m}(\varphi_2),\\
{\hat \sigma_{2}}(U(\Psi_{m}(\varphi_1))) &=_{F}&
{\hat \sigma_{2}}(e^{i\lambda_{+}}\Psi_{m}(\varphi_2))
= e^{i\lambda_{+}}\Psi_{-m}(\varphi_2),
\end{eqnarray*}
and similarly for $m>0$ and $\Psi_{-m}(\varphi_1)
\in {\cal H}_{1}^{-}$. The condition (\ref{32.})
is then equivalent to
\begin{equation}
 e^{i\lambda_{+}}=- e^{i\lambda_{-}}=
          e^{i\lambda}.     \label{34.}
\end{equation}
Thus, the relative phase is determined uniquely
from the condition
\bdm
R_{2}(o)\circ U = U \circ R_{1}(o)
\edm
for any $o$ which maps at least one element of
 ${\cal H}_{1}^{+}$ to ${\cal H}_{1}^{-}$.

The overall phase factor $e^{i\lambda}$ does not
possess any measurable meaning. Indeed, the map
$U_2$ in which it appears is used to carry
quantum mechanical structures (the inner product
and the operators) from ${\cal H}_{1}$ over to
 ${\cal H}$. The result of this transfer is,
however, independent of $\lambda$. Let us choose a
fixed $U_2$; all other possibilities are given by
$e^{i\lambda}U_2$ for $\lambda \in [0,2\pi)$. Then,
two vectors $\Psi_2$ and $\Psi_{2}'$ of ${\cal H}_{2}$
will be mapped to
\bdm
\Psi =   e^{i\lambda}U_{2}\Psi_2 ,\quad
\Psi' =   e^{i\lambda}U_{2}\Psi_{2}' ,
\edm
with
\bdm
(\Psi,\Psi')=(e^{-i\lambda}U_{2}^{-1}\Psi,
 e^{-i\lambda}U_{2}^{-1}\Psi')_{2}=
(U_{2}^{-1}\Psi,U_{2}^{-1}\Psi')_{2}.
\edm
Similarly, if ${\hat O}_2$ is an operator in
 ${\cal H}_{2}$, then its version  ${\hat O}$
in ${\cal H}$ is defined by
\bdm
{\hat O} =  e^{i\lambda}U_{2}{\hat O}_{2}
 e^{-i\lambda}U_{2}^{-1}=U_{2}{\hat O}_{2}U_{2}^{-1}.
\edm
Let us, therefore, set $\lambda =0$ in eq.\ (\ref{34.}).
Then the operator $U_2$ can be expressed in a useful
simple way. Let $\Pi_{2}^{+}$ and  $\Pi_{2}^{-}$
be the projection operators from  ${\cal H}$ onto
 ${\cal H}_{2}^{+}$ and ${\cal H}_{2}^{-}$,
respectively, and similarly  $\Pi^{+}$ and  $\Pi^{-}$
those from ${\cal H}$ onto ${\cal H}^{+}$ and
 ${\cal H}^{-}$. Let further $I:{\cal H}_{2}
\rightarrow {\cal H}$ be defined by
\bdm
I \Psi(\varphi_2) =_{F}\Psi(\varphi).
\edm
Then we have
\begin{eqnarray}
U_2 = I( \Pi_{2}^{+} -\Pi_{2}^{-}),        \label{38.}\\
U_{2}^{-1} = I^{-1}( \Pi^{+} -\Pi^{-}),    \label{39.}
\end{eqnarray}
because of the relations (\ref{35.}), (\ref{34.}) and
(\ref{36.}), (\ref{37.}) (in which $\lambda = 0$).
To check that the right hand side of (\ref{39.}) is
inverse to that of (\ref{38.}), one has to use the
relation
\bdm
I\,\Pi_{2}^{\pm}I^{-1}=\Pi^{\pm}.
\edm
Note that the formulae (\ref{38.}), (\ref{39.})
 enable us to transfer operators and vectors from
  ${\cal H}_{2}$ to  ${\cal H}$ immediately.

\section{Comparison with the Dirac quantization method}

In this section we quantize our system using the straightforward Dirac
quantization method and compare the results with those
 obtained in the preceeding section by reduction method.

Recall that the constraint of our system is
\begin{equation}
C = \frac{1}{2}(p_1^2 - q_{1}^2 - p_2^2 + q_2^2 + \kappa) = 0  \label{b1}
\end{equation}
with $\kappa = 0$.
It is convenient to leave $\kappa$ arbitrary.  We
make the canonical transformation
\bdm
\tilde{q}_1 = p_1,\ \ \ \tilde{p}_1 = -q_1
\edm
and use the Schr\"odinger representation with respect to the variables
$\tilde{q}_1$ and $q_2$.  Thus, the quantum constraint
reads
\begin{equation}
\left( -\frac{\partial^2 \ }{\partial\tilde{q}_1^2}
- \frac{\partial^2 \ }{\partial q_2^2} - \tilde{q}_1^2 - q_2^2 \right) \Psi
= \kappa \Psi. \label{b3}
\end{equation}
The Casimir invariant
$\hat{q} = \hat{L}_0^2 - \hat{L}_1^2 - \hat{L}_2^2$ can be computed using
the canonical
commutation relations and the constraint (\ref{b1}) as
\begin{equation}
\hat{q}\Psi = - \left( \frac{1}{4} + \frac{\kappa^2}{16}\right)\Psi.
\end{equation}
Thus, the value of the Casimir invariant of $SL(2,\reell)$ for $\kappa = 0$
agrees with that found in sec.\ 5.1.

We rewrite (\ref{b3}) using the polar coordinates
\begin{equation}
\tilde{q}_1  = r\cos\phi,\ \ \  q_2 = r\sin\phi
\end{equation}
as
\begin{equation}
\left\{ -\left[ \frac{\partial^2\ }{\partial r^2} + \frac{1}{r}
\frac{\partial\ }{\partial r} + \frac{1}{r^2}
\frac{\partial^2\ }{\partial\phi^2}\right] -r^2\right\}\Psi = \kappa\Psi.
\end{equation}
Note that the angular momentum operator $\hat{L}_0$ is given by
\begin{eqnarray}
\hat{L}_{0} & = & \frac{1}{2}( \tilde{q}_1 p_2 - q_2\tilde{p}_1) \nonumber \\
 & = &
-\frac{i}{2}\frac{\partial\ }{\partial\phi}.
\end{eqnarray}
We choose the wavefunctions to be eigenfunctions of $\hat{L}_0$:
\bdm
\hat{L}_0\Psi^\kappa_m (r, \phi)= \frac{m}{2}\phi, \quad
\Psi_m^{\kappa}(r,\phi) = \psi^{\kappa}_{m} (r) e^{im\phi}.
\edm
 The radial wave equation for $\psi^{\kappa}_m (r)$ is
\begin{equation}
\left[ \frac{d^2\ }{d r^2} + \frac{1}{r}
\frac{d\ }{d r} - \frac{m^2}{r^2} + r^2 + \kappa\right] \psi^{\kappa}_m(r) = 0.
\label{2}
\end{equation}
The solution that is
regular at $r = 0$ is given by
\bdm
\psi_m^{\kappa}(r) \propto \frac{1}{r}
M_{-\frac{i}{4}\kappa,\frac{|m|}{2}}(ir^2),
\edm
where
\bdm
M_{\lambda,\mu}(y) = y^{\mu+\frac{1}{2}}e^{-\frac{y}{2}}
\Phi(\mu -\lambda+\frac{1}{2},2\mu + 1; y)
\edm
is the Whittaker function and
\bdm
\Phi(\alpha,\gamma;y) = 1 + \frac{\alpha}{\gamma}\frac{y}{1!}
+ \frac{\alpha(\alpha + 1)}{\gamma(\gamma+1)}\frac{y^2}{2!} + \cdots
\edm
is the confluent hypergeometric function (see \cite{13},
p.\ 1059, 9.220.1-3).
To find the normalization factor
we need to evaluate
\begin{equation}
\lim_{L \to \infty}\int_{0}^{L}
r\;dr d\phi\; \overline{G_m^{\kappa}(r,\phi)}G_{m'}^{\kappa'}(r,\phi),
\label{integ}
\end{equation}
where
\bdm
G_m^{\kappa}(r,\phi) :=
\frac{1}{r}M_{-\frac{i}{4}\kappa,\frac{|m|}{2}}(ir^2)e^{im\phi}.
\edm
By multiplying by $\kappa - \kappa'$ and using  (\ref{2}), one
can rewrite the integral (\ref{integ}) as a ``surface term" at $r = L$.  Then,
it can be computed using the large $r$ behavior of the wavefunctions.
Using the formula (\cite{13}, p.~1062, 9.233.2)
\begin{eqnarray*}
M_{\lambda,\mu}(y) & = &
\frac{\Gamma(2\mu+1)}{\Gamma(\mu-\lambda+\frac{1}{2})}e^{-i\pi\lambda}
W_{-\lambda,\mu}(e^{-i\pi}y) \nonumber \\
 & & +
\frac{\Gamma(2\mu+1)}{\Gamma(\mu+\lambda+\frac{1}{2})}
\exp\left[ -i\pi\left( \lambda -\mu -\frac{1}{2}\right)\right]
W_{\lambda,\mu}(y)
\end{eqnarray*}
and the fact that the Whittaker function behaves as
$W_{\lambda,\mu}(y)\sim e^{-\frac{y}{2}}y^{\lambda}$
for large $|y|$ with $|{\rm arg}\, y|< \pi$ (\cite{13},
p.~1061, 9.227), we find for large $r$
\begin{equation}
\frac{1}{r}M_{-\frac{i}{4}\kappa,\frac{m}{2}}(ir^2) \sim
\frac{m!e^{-\frac{\pi\kappa}{8}}}{\Gamma(\frac{m+1}{2}+\frac{i\kappa}{4})}
e^{ir^2}r^{i\frac{\kappa}{2}-1}
+ \frac{m!e^{\frac{m+1}{2}\pi i}e^{-\frac{\pi\kappa}{8}}}
{\Gamma(\frac{m+1}{2}-\frac{i\kappa}{4})}
e^{-ir^2}r^{-i\frac{\kappa}{2}-1}.
\label{3}
\end{equation}
By using this equation,
dropping rapidly oscillating functions of $\kappa$ and $\kappa'$ and using
\bdm
\lim_{\alpha \to +\infty}\frac{\sin\alpha x}{x} = \pi\delta(x),
\edm
we find
\bdm
\lim_{L\to \infty}\int_0^{L} r\; dr\; d\phi\;
\overline{G_{m}^{\kappa}(r,\phi)}G_{m'}^{\kappa'}(r,\phi)
 =  \frac{8\pi^2(|m|!)^2e^{-\frac{\pi\kappa}{4}}}
{\left| \Gamma\left(\frac{|m|+1}{2}+\frac{i}{4}\kappa\right)\right|^2}
\delta_{mm'}\delta(\kappa-\kappa').
\edm
It is clear that this integral is divergent if we let $\kappa = \kappa' = 0$.
This divergence simply represents the volume of the ``gauge group'' generated
by the constraint. This implies that the inner product of wavefunctions
satisfying the constraint should be defined by dropping
$2\pi\delta(0)$,
as recently proposed by Marolf \cite{14}.\footnote{
This procedure is closely related to
that introduced to deal with ``linearization instabilities''
 \cite{15}.}
Hence we define
\begin{equation}
\Psi_m^{\kappa}(r,\phi) :=
\frac{\Gamma\left(\frac{|m|+1}{2}+\frac{i}{4}\kappa\right)}
{\sqrt{4\pi}|m|!}\times
\frac{1}{r}M_{-\frac{i}{4}\kappa,\frac{|m|}{2}}(ir^2)e^{im\phi}. \label{def}
\end{equation}
They satisfy
\bdm
\int r\; dr\;d\phi\;
\overline{\Psi_m^{\kappa}(r,\phi)}\Psi_{m'}^{\kappa'}(r,\phi) =
2\pi e^{-\frac{\pi\kappa}{4}}\delta(\kappa-\kappa')\delta_{mm'}.
\edm
Then the normalized basis functions satisfying the constraint
 (\ref{b3}) with $\kappa =0$ are given by
\bdm
\Psi_m(r,\phi) = \Psi^0_m(r,\phi).
\edm
The inner product is defined simply by
\bdm
\langle \Psi_m\vert \Psi_{m'}\rangle = \delta_{mm'}.
\edm

Now, notice that $r^m e^{im\phi} = (\tilde{q}_1 + i q_2)^m$.  By defining
\begin{eqnarray*}
z & := & \frac{\tilde{q}_1 + iq_2}{2}, \\
\bar{z} & := & \frac{\tilde{q}_1 - iq_2}{2},
\end{eqnarray*}
and using the expression  (\ref{def}) for
 $\Psi_m^{\kappa}(r,\phi)$ we find
\begin{equation}
\Psi_{m}^{\kappa} = \frac{i^{\frac{m+1}{2}}e^{-2i\bar{z}{z}}}
{\sqrt{4\pi}}\sum_{k=0}^{\infty}
\frac{\Gamma(\frac{m+1}{2}+k+\frac{i\kappa}{4})}{(m+k)!k!}i^k
(2\bar{z})^k (2z)^{k+m}\ \ (m\geq 0)
\label{4}
\end{equation}
and
\begin{equation}
\Psi_{m}^{\kappa} = \frac{i^{\frac{|m|+1}{2}}e^{-2i\bar{z}{z}}}
{\sqrt{4\pi}}\sum_{k=0}^{\infty}
\frac{\Gamma(\frac{|m|+1}{2}+k+\frac{i\kappa}{4})}{(|m|+k)!k!}i^k
(2z)^k (2\bar{z})^{k+|m|}\ \ (m < 0).  \label{44}
\end{equation}
It is straightforward to derive the following formulae:
\begin{eqnarray}
\left(\frac{1}{2}\frac{\partial\ }{\partial z} +
i\bar{z}\right)\Psi_m^{\kappa} & = & i^{\frac{1}{2}}\Psi_{m-1}^{\kappa -2i},
\label{ba} \\
\left(\frac{1}{2}\frac{\partial\ }{\partial \bar{z}} +
iz\right)\Psi_m^{\kappa} & = & i^{\frac{1}{2}}\Psi_{m+1}^{\kappa -2i},
\label{bb} \\
\left(\frac{1}{2}\frac{\partial\ }{\partial z} -
i\bar{z}\right)\Psi_m^{\kappa} & = & \left(\frac{m-1}{2} +
\frac{i}{4}\kappa\right) i^{\frac{1}{2}}\Psi_{m-1}^{\kappa +2i}, \label{bc}\\
\left(\frac{1}{2}\frac{\partial\ }{\partial \bar{z}} +
iz\right)\Psi_m^{\kappa} & = & \left( - \frac{m+1}{2} + \frac{i}{4}\kappa
\right) i^{\frac{1}{2}}\Psi_{m+1}^{\kappa +2i}. \label{bd}
\end{eqnarray}
Noting that
\begin{eqnarray*}
\tilde{q}_1  & =  & z + \bar{z}, \\
q_2 & = & i(\bar{z} - z), \\
\tilde{p}_1 & = & -\frac{i}{2}\left(\frac{\partial\ }{\partial z}
+ \frac{\partial\ }{\partial\bar{z}}\right), \\
p_2 & = & \frac{1}{2}\left(\frac{\partial\ }{\partial z}
- \frac{\partial\ }{\partial\bar{z}}\right),
\end{eqnarray*}
we find
\begin{eqnarray*}
\hat{L}_{+} & = & -i\left( \frac{1}{4}\frac{\partial^2\ }{\partial\bar{z}^2}
+ z^2 \right),\\
\hat{L}_{-} & = & i \left( \frac{1}{4}\frac{\partial^2\ }{\partial z^2}
+ \bar{z}^2\right),
\end{eqnarray*}
where $\hat{L}_{\pm}$ are defined by (\ref{42.}).
It is easy to find how these operators act on $\Psi_m^{\kappa}$ using
(\ref{ba})--(\ref{bd}):
\begin{eqnarray*}
\hat{L}_{+}\Psi_{m}^{\kappa} & = & \left(
-\frac{m+1}{2}+\frac{i\kappa}{4}\right)
\Psi_{m+2}^{\kappa},\\
\hat{L}_{-}\Psi_{m}^{\kappa}& = & \left( - \frac{m-1}{2} -
\frac{i\kappa}{4}\right) \Psi_{m-2}^{\kappa},
\end{eqnarray*}
in agreement with relation (\ref{43.}) obtained in sec.\ 5.1.

Next, we consider the operators $\hat{\psi}_1$ and $\hat{\psi}_2$
whose classical counterparts are defined in sec.\ 2.3.  By
expressing the operators $\hat{Q}_1$, $\hat{Q}_2$, $\hat{P}_1$,
and $\hat{P}_2$
in terms of $z$, $\bar{z}$, $\partial/\partial z$, and
$\partial/\partial\bar{z}$, one finds
\begin{eqnarray*}
e^{i\hat{\psi}_1} & = &
\left( \frac{1}{2}\frac{\partial\ }{\partial\bar{z}} + iz
\right) \left( \frac{1}{2}\frac{\partial\ }{\partial z} +
i\bar{z}\right)^{-1}, \\
 e^{i\hat{\psi}_2} & = &
-\left( \frac{1}{2}\frac{\partial\ }{\partial\bar{z}} - iz
\right) \left( \frac{1}{2}\frac{\partial\ }{\partial z} - i\bar{z}\right)^{-1}
\end{eqnarray*}
Notice that there is no factor ordering ambiguity.  However, these
operators are ill-defined for $\kappa = 0$.  For this reason we regularize
them by allowing $\kappa$ to be nonzero, and then take the
limit $\kappa \to 0$.
We find using (\ref{ba})--(\ref{bd})
\begin{eqnarray*}
e^{i\hat{\psi}_1}\Psi_m^{\kappa} & = & \Psi_{m+2}^{\kappa}, \\
e^{i\hat{\psi}_2}\Psi_m^{\kappa}  &= & \frac{m+1-\frac{i}{2}\kappa}{m+1 +
\frac{i}{2}\kappa} \Psi_{m+2}^{\kappa},
\end{eqnarray*}
By taking the limit $\kappa \to 0$, we have
\begin{eqnarray*}
e^{i\hat{\psi}_1}\Psi_m = \Psi_{m+2}\ \ ({\rm for\ all\ }m).
\end{eqnarray*}
and
\begin{eqnarray*}
e^{i\hat{\psi}_2}\Psi_m & = & \Psi_{m+2}\ \ (m \neq -1), \\
e^{i\hat{\psi}_2}\Psi_{-1} & = & -\Psi_{+1},
\end{eqnarray*}
In
sec.\ 5, quantization was performed in the representations
in which $\hat{\psi}_1$ or $\hat{\psi}_2$ is diagonalized.
The wavefunction in our Schr\"odinger representation,
\bdm
\Phi(r,\phi) = \sum_{m=-\infty}^{+\infty}c_m\Psi_m(r,\phi),
\edm
is written in those representations as
\begin{eqnarray}
\Psi^{(1)}(\varphi_1) & = & \sum_{m=-\infty}^{+\infty}
c_m e^{i\frac{m}{2}\varphi_1}, \label{11} \\
\Psi^{(2)}(\varphi_2) & = &
\sum_{m=+1}^{+\infty} c_m e^{i\frac{m}{2}\varphi_2}
- \sum_{m = -\infty}^{-1} c_m e^{i\frac{m}{2}\varphi_2}, \label{10}
\end{eqnarray}
where the summation is over odd $m$. We have identified $\psi_\lambda$ with
$\varphi_\lambda (\lambda = 1,2)$ since the variable $e^{i\psi_\lambda}$ is
naturally associated with $\Gamma_\lambda$
[see Sec.\ 4, below (\ref{133})]. The
maps $\Phi \to \Psi^{(i)}$ are
unitary with the inner products of $\Psi^{(i)}$ defined as in (\ref{140}).
These wavefunctions indeed satisfy
\begin{eqnarray*}
e^{i\hat{\psi}_i}\Psi(\varphi_i) & = &
e^{i\varphi_i}\Psi^{(i)}(\varphi_i), \\
\hat{L}_0 \Psi^{(i)}(\varphi_i) & = & -i\frac{\partial\ }{\partial\varphi_i}
\Psi^{(i)}(\varphi_i), \\
\hat{L}_{\pm}\Psi^{(i)}(\varphi_i) & = & i e^{\pm i\frac{\varphi_i}{2}}
\frac{\partial\ }{\partial\varphi_i}e^{\pm
i\frac{\varphi_i}{2}}\Psi^{(i)}(\varphi_i).
\end{eqnarray*}

Finally, we consider the discrete operator $\hat{\sigma}$
corresponding to the classical transformation defined in sec.\ 3.2.
The operator $\hat{\sigma}$ multiplies $\tilde{q}_1$ and
$\tilde{p}_1 = -i\partial/\partial \tilde{q}_1$ by $-1$
in the Dirac quantization adopted
here.  Hence,
\bdm
\hat{\sigma} z = -\bar{z}, \ \ \ \hat{\sigma} \bar{z} = -z.
\edm
Using (\ref{4}) and (\ref{44}) we find
\bdm
\hat{\sigma} \Psi_m = -\Psi_{-m}.
\edm
(Recall that we have restricted $m$ to be odd.)  Thus, from (\ref{10}) and
(\ref{11}) the corresponding operator $\hat{\sigma}$ on $\Psi^{(1)}$ and
$\Psi^{(2)}$ can be found as
\begin{eqnarray*}
\hat{\sigma}\Psi^{(1)}(\varphi_1) & = & \Psi^{(1)}(2\pi -\varphi_1), \quad
\varphi \in [0,2 \pi),\\
& = & \Psi^{(1)} (6 \pi - \varphi_1), \quad \varphi_1 \in [2 \pi, 4 \pi) \\
\hat{\sigma}\Psi^{(2)}(\varphi_2) & = &
\Psi^{(2)}(4\pi-\varphi_2),\quad \varphi_2 \in [0, 4\pi).
 \end{eqnarray*}
It is interesting to note that these formulae coincide with
eqs.\ (\ref{44.}) --
(\ref{45.}).

\section{Time levels}

Consider the transversal surface $\Gamma_i$. The subgroup of $O^+_2(2,1)$ that
leaves $\Gamma_i$ invariant is generated by $\Lambda_0(t)$, $t \in \reell$,
and $\sigma$. All other generators of $SO_2(2,1)$ are, therefore, possible
candidates for Hamiltonians.

The formulae (22), (23), (29), (30) and (31) show how the subgroup
$\Lambda_\theta (t)$ shifts the surface $\Gamma_i$. Let us denote
$\Gamma_i(t,\theta)$
the result; then, $\Gamma_i(t,\theta)$ is a two-dimensional family of time
levels. We will demonstrate that this is a complete family.

The intersection of $\Gamma_i (t,\theta)$ with the surface $\psi =$ const is a
curve
\begin{displaymath}
x= TN_\theta (\psi,t)
\end{displaymath}
for $i = 1$ and
\begin{displaymath}
y=TN_\theta (\psi, t)
\end{displaymath}
for $i=2$. If we fix $\theta$ and $\psi$, then the function $N_\theta(\psi, t)$
has a minimum at $t_m(\theta, \psi)$,
\begin{displaymath}
t_m (\theta, \psi)\; = \; \mbox{arctanh} \sin (\psi + \theta)
\end{displaymath}
given by
\begin{displaymath}
N_\theta (\psi, t_m (\theta, \psi)) = \sqrt{\vert \cos (\psi + \theta)\vert}.
\end{displaymath}
The minimum is {\em positive} unless
$\psi \in \left \{ \frac{\pi}{2} - \theta, \frac{3\pi}{2} - \theta, \frac{5
\pi}{2} - \theta, \frac{7 \pi}{2} - \theta \right \}$, where $t_m (\theta,
\psi) = \pm \infty$ and $ N_\theta (\psi, t_m (\theta, \psi)) = 0$. Thus, the
surfaces $\Gamma_i(t, \theta)$ for a fixed $\theta$ do not sweep the whole
domain of $\Gamma_i$ but only a part satisfying
\bdm
x \geq T \sqrt{\vert \cos (\psi + \theta) \vert}
\edm
for $i = 1$ and
\bdm
y \geq T \sqrt{\vert \cos (\psi + \theta) \vert}
\edm
for $i = 2$. Hence, each $L_\theta$ can only be made a partial Hamiltonian.
However, the set of $\Gamma_i (t, \theta)$ for all $t$ and $\theta$ cover the
whole $\mbox{D} (\Gamma_i)$. Indeed,
\bdm
\mbox{D} (\Gamma_1) = \{(x,y, \psi) \in \tilde{\Gamma}\; \vert\;
x > 0, - \infty <
y < \infty, 0 < \psi \leq 4\pi\}
\edm
\bdm
\mbox{D}(\Gamma_2) = \{(x,y, \psi) \in \tilde{\Gamma}\; \vert\;
y > 0, - \infty <
x < \infty, 0 < \psi \leq 4 \pi \}.
\edm
Thus, the one-dimensional family of generators $L_\theta$, $\theta \in (0, 2
\pi)$,
yields a complete system of partial Hamiltonians for each of the surfaces
$\Gamma_1$ and $\Gamma_2$. We obtain two disjoint systems of time levels (no
element of our first-class canonical group maps a time level of one system into
a time level of the second system). Let us limit ourselves to $\Gamma_1(t,
\theta)$, and consider the problem of the ``same measurement'' at different
time
level. (We leave out the index ``1'' referring to the surface $\Gamma_1$ in the
rest of this section. For example, $\varphi_1$ is denoted simply by $\varphi$.)

According to Theorem 9 in I, we can calculate the change in the result of the
``same measurement'' represented by a perennial $o$ between two neighbouring
levels $\Gamma_1(t, \theta)$ and $\Gamma_1(t+dt, \theta)$ as follows. We
restrict $o$ to $\Gamma_1$ obtaining the time-independent observable
$\tilde{o}$ and calculate its value along the ``physical trajectory''
generated in $\Gamma_1$ by the ``Hamiltonian'' $ - L_\theta\vert_{\Gamma_1}$.
This is the Schr\"{o}dinger picture. To be more concrete, we have the
observables given by eqs.\ (38) and (39) with $i = 1$, and we have the
Hamiltonians
\bdm
H_\theta = L_\theta \vert_{\Gamma_1} = - L_1 \cos \theta - L_2 \sin \theta = l
\cos (\varphi + \theta).
\edm
In the quantum theory, the observables are represented by the time-independent
operators $\hat{L}_0, \hat{L}_1, \hat{L}_2$ and $\hat{\sigma}$ given by
eqs.\ (41)--(43) and (46)--(48), whereas the Hamiltonian operator
$\hat{H}_\theta$ can be calculated with the result
\bdm
\hat{H}_\theta = - i \cos (\varphi + \theta) \frac{\partial}{\partial \varphi}
+ \frac{i}{2} \sin (\varphi + \theta).
\edm
This is, of course, $- \hat{L}_\theta$ as given by eq.\ (\ref{145}).
The corresponding
Schr\"{o}dinger equation
\bdm
i \frac{\partial \Psi}{\partial t} = \hat{H}_\theta \Psi
\edm
is then easily integrated, if we use eqs.\ (44) and (45), with the result
\bdm
\Psi (\theta, t, \varphi) = (U (\Lambda_\theta (t)) \Psi) (\varphi)
= \frac{1}{N_\theta(\varphi, t)} \Psi (F_\theta(\varphi, t)).
\edm
Thus, we have a four-dimensional family of unitary operators, $U^{-1}
(\Lambda_{\theta'} (t')) U (\Lambda_{\theta} (t))$, that bring the
wave function $\Psi (\theta', t', \varphi)$ at the time level
$\Gamma_1(\theta',t')$ to the wave function $\Psi (\theta, t, \varphi)$ at the
time level $\Gamma_1 (\theta,t)$. The corresponding Heisenberg picture is
easy to obtain, and we leave it to the reader as an exercise.

An important observation is the following. For the system we are studying in
the present paper, the full Hilbert space ${\cal H}$ is identical to the
Hilbert
space ${\cal H}_1$ corresponding to the reduced subsystem $(\Gamma_1,
\Omega_1)$.
Hence, our operators that represent both observables and Hamiltonians are
applicable to {\em all} vectors of ${\cal H}$. Hence, we have already a
complete quantum theory and a complete ``truncated'' many-fingered time
evolution.
Clearly, the set of classical orbits of measure zero that is missing in
$(\Gamma_1, \Omega_1)$ (namely orbits with $x = 0$) does not play any role in
the quantum theory. If one would like to study wave packets concentrated along
these orbits, then one clearly must use the same Hilbert space ${\cal H}$, but
now ``in the form" of ${\cal H}_2$ (which is unitarily equivalent).

To summarize: we have found a complete two-dimensional family of time levels at
the constraint hypersurface of the system and we have constructed the
four-dimensional family of unitary operators representing the corresponding
(``truncated many-fingered'') time evolution.

\subsection*{Acknowledgements}
We are grateful for discussions with J. Louko, D. Marolf, J. Mour\~{a}o,
and R. Tate. P.H. and A.H. acknowledges the support of the Swiss Nationalfonds.
J.T. is indebted for
the warm hospitality of the Institute for Theoretical Physics in Bern. He also
 acknowledges the support of the Tomalla-Stiftung
 Fund and of the Grant Agency of Czech Republic.

\end{document}